\documentclass[a4paper,11pt]{article}
\pdfoutput=1
\usepackage{jheppub}
\usepackage{amsmath}
\usepackage{hyperref}
\allowdisplaybreaks
\def\bea#1\eea{\begin{align}#1\end{align}}
\def \be  {\begin{equation}}
\def \ee  {\end{equation}}
\newcommand{\nnu}{\nonumber\\}

\newcommand{\bef}{\begin{figure}[t]\centering}
\newcommand{\eef}{\end{figure}}
\newcommand{\sla}[1]{{#1}\!\!\!\slash}

\newcommand{\f}{\frac}

\title{The groomed and ungroomed jet mass distribution for inclusive jet production at the LHC}

\author[a,b,c]{Zhong-Bo Kang,}
\author[d,e]{Kyle Lee,}
\author[f]{Xiaohui Liu}
\author[g]{and Felix Ringer}

\affiliation[a]{Department of Physics and Astronomy, University of California, Los Angeles, CA 90095, USA}
\affiliation[b]{Mani L. Bhaumik Institute for Theoretical Physics, University of California, Los Angeles, CA 90095, USA}
\affiliation[c]{Theoretical Division, Los Alamos National Laboratory, Los Alamos, NM 87545, USA}
\affiliation[d]{C.N. Yang Institute for Theoretical Physics, Stony Brook University, Stony Brook, NY 11794, USA}
\affiliation[e]{Department of Physics and Astronomy, Stony Brook University, Stony Brook, NY 11794, USA}
\affiliation[f]{Center of Advanced Quantum Studies, Department of Physics, Beijing Normal University, Beijing 100875, China}
\affiliation[g]{Nuclear Science Division, Lawrence Berkeley National Laboratory, Berkeley, CA 94720, USA}
                                   
\emailAdd{zkang@physics.ucla.edu}
\emailAdd{kunsu.lee@stonybrook.edu}
\emailAdd{xiliu@bnu.edu.cn}
\emailAdd{fmringer@lbl.gov}

\abstract{We study jet mass distributions measured in the single inclusive jet production in proton-proton collisions $pp\to \text{jet}+X$ at the LHC. We consider both standard ungroomed jets as well as soft drop groomed jets. Within the Soft Collinear Effective Theory (SCET), we establish QCD factorization theorems for both cases and we study their relation. The developed framework allows for the joint resummation of several classes of logarithmic corrections to all orders in the strong coupling constant. For the ungroomed case, we resum logarithms in the jet radius parameter and in the small jet mass. For the groomed case, we resum in addition the logarithms in the soft threshold parameter $z_{\text{cut}}$ which is introduced by the soft drop grooming algorithm. In this way, we are able to reliably determine the absolute normalization of the groomed jet mass distribution in proton-proton collisions. All logarithmic corrections are resummed to the next-to-leading logarithmic accuracy. We present numerical results and compare with the available data from the LHC. For both the groomed and ungroomed jet mass distributions we find very good agreement after including non-perturbative corrections.}

\begin{document}
\maketitle

\section{Introduction \label{sec:intro}}
In high energy collisions the fundamental building blocks of Quantum Chromodynamics (QCD), quarks and gluons, lead to the formation of highly energetic collimated sprays of hadrons observed in the detectors which are known as jets~\cite{Sterman:1977wj}. The Large Hadron Collider (LHC) is currently the world's largest and highest energy particle collider where jets provide a unique opportunity to test the fundamental properties of QCD and to search for new physics beyond the standard model at the TeV scale. Therefore jet studies have become one of the most important topics both in the experimental and theoretical communities. One of the most studied benchmark processes at the LHC is the inclusive jet production cross section differential in the jet rapidity and the transverse momentum~\cite{Aaboud:2017dvo,Khachatryan:2016mlc,Abelev:2013fn}. Over the past years, it has been realized that the internal structure of the identified jets contains additional valuable information. When additional measurements are performed on the identified jets in order to characterize and utilize the radiation pattern inside jets, the corresponding observables are generally referred to as jet substructure measurements~\cite{Larkoski:2017jix}. For example, jet substructure techniques are used to improve our understanding of the QCD hadronization mechanism~\cite{Chatrchyan:2012gw,Aad:2011td,Aaij:2017fak}, to discriminate between quark and gluon jets~\cite{Gras:2017jty} or to identify jets originating from the decay of boosted objects~\cite{Abdesselam:2010pt}. At the LHC heavy particles such as $W/Z$, Higgs, and top quarks are often produced with a high transverse momentum such that their decay products become collimated and thus are merged into a single jet. The radiation pattern of the produced jets contains information about the different decaying particles. In order to tag such boosted objects and to separate them from the QCD background, jet substructure techniques have proven to be an invaluable tool. In addition, jet substructure techniques are used increasingly for the search of new resonances from physics beyond the standard model. See for example~\cite{Sirunyan:2017dnz} for a recent search for hadronically decaying vector resonances reported by the CMS collaboration relying on jet substructure techniques. Often several jet substructure observables are measured on a single jet in order to enhance the tagging efficiency, see for example~\cite{Moult:2017okx}. In addition, jet substructure observables are increasingly being studied in heavy-ion collisions where they provide an important test of the hot and dense QCD medium~\cite{Wang:2016opj}.

One of the most prominent and most often used jet substructure observables is the jet mass distribution which we address in this work in the context of inclusive jet production in proton-proton collisions at the LHC. We consider the cross section where the jet mass is measured for jets that are identified with a given transverse momentum $p_T$ and rapidity $\eta$ using a jet radius parameter $R$ and where the measurement is inclusive about everything else in the event which is denoted by $X$. Thus, we have 
\bea
pp\to {\rm jet}(\tau; \eta, p_T, R) + X, 
\eea
where we introduced the dimensionless variable $\tau$ which is related to the jet invariant mass $m_J$ as
\bea
m_J^2 = \Big(\sum_{i\in J} p_i\Big)^2\,,
\qquad
\tau = \frac{m_J^2}{p_T^2} \,.
\eea
Here, $p_i$ are the four-momenta of all the particles inside the reconstructed jet. More specifically we consider the normalized jet mass distribution
\bea
\label{obs:ungroom}
F(\tau;\eta,p_T,R) = \left.\frac{d\sigma}{d\eta dp_T d\tau}  \middle/ \f{d\sigma}{d\eta dp_T}\right. \,,
\eea
where the numerator and the denominator are the differential jet cross sections with and without the additional measurement of the jet mass, respectively. Traditionally, jet mass measurements have been performed on an inclusive jet sample, see for example the data sets in $p\bar p$~\cite{Aaltonen:2011pg} and $pp$ collisions~\cite{ATLAS:2012am} by the CDF collaboration at Tevatron and the ATLAS collaboration at the LHC, respectively. In addition, inclusive jet mass measurements have been performed by the ALICE collaboration in Pb-Pb and p-Pb collisions at the LHC~\cite{Acharya:2017goa}.

Although being one of the simplest and most intuitive examples of jet substructure observables, the jet invariant mass spectrum serves as a benchmark observable for jet substructure studies and is therefore of great phenomenological relevance. The jet mass distribution is used to test parton showers in Monte Carlo event generators, to tag quark-gluon jets, and to search for boosted objects as outlined above. In addition, it is expected that jet mass measurements can shed new light on the jet quenching phenomenon observed in heavy-ion collisions. Even though jet mass measurements are of great phenomenological importance, current studies of the inclusive jet mass spectrum rely heavily on the assumption that the jet mass distribution is well modeled by Monte Carlo event generators. Recent studies by the ATLAS and ALICE collaborations suggest that this assumption should be treated with care. For instance, ATLAS found that the predicted spectrum by Pythia~\cite{Sjostrand:2006za} is too soft in $pp$ collisions whereas the one from Herwig++~\cite{Bahr:2008pv} is too hard~\cite{ATLAS:2012am}. A similar situation was observed in the heavy-ion collisions, where the jet mass distribution is over- or underestimated by Q-Pythia~\cite{Armesto:2009fj} and Jewel~\cite{Zapp:2012ak} depending on the out-of-jet radiation settings~\cite{Acharya:2017goa}. Therefore, jet mass calculations from first principles in QCD are needed in order to improve our understanding of the underlying mechanisms and to benchmark current models.

The jet mass distribution has been addressed several times in the literature from the theoretical side, where the efforts have focused mostly on exclusive jet configurations where additional constraints are imposed on the final state particles, see for example~\cite{Jouttenus:2013hs,Stewart:2014nna,Hornig:2016ahz,Kolodrubetz:2016dzb} and references therein. While it is advantageous in some situations to consider exclusive final state jets, it is important to note that the inclusive jet cross section can be measured with the highest statistics since all jets in a given transverse momentum and rapidity interval are taken into account without any further restrictions. So far only a few theoretical studies exist in the literature on the inclusive jet mass spectrum~\cite{Li:2011hy,Li:2012bw,Liu:2014oog,Idilbi:2016hoa}. In \cite{Liu:2014oog}, the jet mass distribution was calculated in the threshold di-jet limit, while \cite{Li:2011hy,Li:2012bw} focused on process-independent jet functions. See also the theoretical studies in~\cite{Dasgupta:2012hg,Chien:2012ur} on the jet mass distributions in $\gamma/Z$+jet and di-jet processes at the LHC, as well as the experimental measurements at the LHC~\cite{Chatrchyan:2013vbb}. In this work, we derive a complete factorization theorem from first principles in QCD allowing for a direct comparison with the inclusive jet mass data from the LHC. Using the QCD factorization theorem derived in this work, we are able to jointly resum single logarithms in the jet size parameter $R$ and double logarithms in the jet mass $m_J$.

As it turns out, the invariant jet mass distribution is very sensitive to the soft hadronic activity in the collisions recorded at the LHC. This is the case in particular at the highest energy collisions currently achieved at the LHC at $\sqrt{s}=13$~TeV. The soft radiation includes pileup and the underlying event contribution like multi-parton interactions (MPI)~\cite{Soyez:2018opl}. The jet mass distribution as introduced above may, in fact, play an important role in order to disentangle the various contributions, see for example~\cite{Stewart:2014nna}. However, for many applications, it is important to remove the underlying event contribution in order to restore the understanding of the jet mass as being a direct measure of the mass associated with a highly energetic fragmenting parton or a boosted object that produces the observed final state jet. This can be achieved by considering the groomed jet mass which we denote by $m_{J,{\rm gr}}$. Various jet grooming techniques have been developed in the past decade, see for example~\cite{Larkoski:2017jix} for an overview. The grooming procedure which we use in this work is the so-called soft drop grooming algorithm~\cite{Larkoski:2014wba} which can be included in analytical calculations using QCD factorization theorems. The soft drop grooming algorithm is designed to remove wide-angle soft radiation from the jet by recursively declustering a jet and by removing soft branches from the identified initial ungroomed jet. The algorithm depends on the angular separation of the branches obtained at each declustering step, an angular exponent $\beta$, and a soft threshold $z_{\text{cut}}$ which sets the cutoff below which soft branches are removed from the jet. See section~\ref{sec:softdrop} for a more detailed description of the soft drop grooming algorithm. After all the soft branches of a given jet have been identified and removed from the jet any observables may be measured on the remaining jet constituents~\cite{Larkoski:2017bvj,Larkoski:2017cqq,Makris:2017arq}. An important feature of the grooming procedure is that it reduces the sensitivity to non-perturbative contributions and non-global logarithms (NGLs)~\cite{Dasgupta:2001sh,Banfi:2002hw} which we discuss in more detail below. 

In this work, we focus on the soft drop groomed jet mass distribution which we are going to compare to the ungroomed case as discussed above. Earlier work on groomed jet mass  distributions can be found in~\cite{Frye:2016aiz,Marzani:2017mva,Marzani:2017kqd}. Here, we derive a complete QCD factorization theorem that allows for the resummation of three important logarithmic corrections to all orders in the strong coupling constant $\alpha_s$: Single logarithms in the jet size parameter $R$ and double logarithms in the jet mass $m_{J,{\rm gr}}$ similar to the ungroomed case. In addition, we are able to completely resum logarithms in the soft threshold parameter $z_{\text{cut}}$ which was not achieved for jets in $pp$ collisions before. Using the new framework developed for inclusive jet samples it is therefore possible to reliably determine for the first time the absolute normalization of a groomed jet observable up to NGLs. All resummations in this work are carried out at next-to-leading logarithmic accuracy. Throughout this paper, we derive QCD factorization theorems and the resummation of large logarithms within the framework of Soft Collinear Effective Theory (SCET)~\cite{Bauer:2000ew,Bauer:2000yr,Bauer:2001ct,Bauer:2001yt,Bauer:2002nz}. The experimental measurements of the jet mass distribution for soft drop groomed jets in $pp$ collisions have been performed by both CMS~\cite{Chatrchyan:2013vbb,CMS:2017tdn} and ATLAS~\cite{Aaboud:2017qwh} collaborations at the LHC. 

Recently there has been a great interest in groomed jet observables in the heavy-ion community~\cite{Sirunyan:2017bsd,Kauder:2017cvz}. It is generally desirable to use grooming in order to directly probe the medium modification of highly energetic fragmenting partons that produce a jet in the final state which traverses and thus probes the QCD medium. On the other hand, one has to worry about the interference of the grooming procedure with the employed background subtraction method. In order to disentangle the interplay of the grooming procedure and the subtraction of the background, it is absolutely crucial to consider observables that are defined both with and without grooming. The jet mass distributions discussed in this work constitute an ideal starting point for further studies along these lines as they allow for a continuous transition between the groomed and ungroomed case. See for example~\cite{Sirunyan:2018gct}, where the CMS collaboration reported on the results for the groomed jet mass distribution in Pb-Pb collisions at the LHC.

The remainder of this paper is organized as follows. In Sec.~\ref{sec:ungroom-fac} we derive the factorization for the ungroomed jet mass distribution. In Sec.~\ref{sec:groom-fac} we extend the obtained framework to include soft drop grooming. We emphasize similarities and differences between the groomed and ungroomed jet mass distribution. In Sec.~\ref{sec:NGLs} we briefly comment on the different NGLs that contribute to the groomed and the ungroomed jet mass spectrum and we comment on the relation of our newly derived factorization to earlier work in the literature. In Sec.~\ref{sec:numerics} we present numerical results for both jet mass distributions and we compare to the available experimental results from the LHC. We summarize our results in Sec.~\ref{sec:summary} and present an outlook. Several detailed calculations of the relevant soft functions at one-loop order in the presence of soft drop grooming are presented in the Appendix~\ref{sec:appendixA}. 

\section{Factorization: the ungroomed jet mass \label{sec:ungroom-fac}}
In this section, we present the factorization formalism of the ungroomed jet mass distribution for single inclusive jet production. We closely follow the arguments provided recently in~\cite{Kang:2018qra} where jet angularities were discussed for inclusive jets. The jet mass distribution is a special case of the jet angularity observables $\tau_a$ with $a=0$. See~\cite{Kang:2018qra} for more detailed discussions. In the small jet mass limit, the factorization procedure involves two steps. The first step is a hard collinear factorization, which describes the production of a single inclusive jet with radius $R$. The second step deals with the factorization of the details of the jet substructure measurement~(i.e. the jet mass $m_J$ or $\tau$) in terms of soft and collinear modes. 

\subsection{First step: hard collinear factorization}
For the jet mass distribution measured in the single inclusive jet production in $pp$ collisions, the factorized cross section in the small jet radius limit can be written as
\bea
\label{eq:fac-ungroom}
 \f{d\sigma}{d\eta dp_T d\tau} = \sum_{abc} f_a(x_a,\mu) \otimes f_b(x_b,\mu) \otimes H_{ab}^c(x_a,x_b,\eta,p_T/z,\mu)\otimes {\cal G}_c(z, p_T,R, \tau,\mu)\;,
\eea
where $f_{a,b}$ denote the parton distribution functions (PDFs) of the colliding protons with momentum fractions $x_{a,b}$. The hard functions $H_{ab}^c$ describe the production of an energetic parton $c$ in the hard-scattering event similar to the inclusive production of hadrons~\cite{Aversa:1988vb,Jager:2002xm}. In fact, it was shown in~\cite{Kang:2016mcy} that the hard functions are exactly the same as those for the single inclusive hadron production, $pp\to h+X$. The functions ${\cal G}_c(z,p_T, R,\tau,\mu)$ are the semi-inclusive jet mass functions (siJMFs), which describe how a parton $c$ initiates the signal jet that carries a momentum fraction $z$ of that parton, and at the same time the jet mass $\tau$ is observed. Following earlier work~\cite{Kang:2016ehg,Kang:2017yde,Kang:2017glf,Kang:2018qra,Kaufmann:2015hma}, the siJMFs are defined at the operator level as follows
\bea
{\cal G}_q(z,p_T, R,\tau,\mu) &= \f{z}{2N_c}{\rm Tr} \left[\f{\sla{\bar n}}{2}
\langle 0| \delta\left(\omega - \bar n\cdot {\mathcal P} \right)\delta(\tau - \hat \tau(J)) \chi_n(0)  |JX\rangle \langle JX|\bar \chi_n(0) |0\rangle \right]\,,
\\
{\cal G}_g(z,p_T, R,\tau,\mu) &= - \f{z\,\omega}{2(N_c^2-1)}
\langle 0|  \delta\left(\omega - \bar n\cdot {\mathcal P} \right)\delta(\tau - \hat \tau(J)) {\mathcal B}_{n\perp \mu}(0) 
 |JX\rangle \langle JX|{\mathcal B}_{n\perp}^\mu(0)  |0\rangle\,.
\eea
Here $\chi_n$ and ${\mathcal B}_{n\perp}^\mu$ are the SCET gauge invariant collinear quark and gluon fields, and ${\cal P}$ is the label momentum operator. Here, we have defined two light-like vectors $n^\mu = (1, \hat n)$ and $\bar n^\mu =(1, -\hat n)$ with $n^2=\bar n^2 =0$ and $n\cdot \bar n = 2$, where $\hat n$ is aligned along the jet axis. The operator $\hat \tau(J)$ represents the jet mass measurement for the observed jet $J$, with the measured value being equal to $\tau$. 

This first step of the factorization of the siJMFs ${\cal G}_c$ from the hard functions $H_{ab}^c$ is the so-called hard collinear factorization~\cite{Bauer:2002nz}, which specifically describes the production of a jet with rapidity $\eta$, transverse momentum $p_T$, and jet radius $R$. To derive this factorization, we work with parametrically small values of the jet size parameter $R\ll 1$. In this case, we have two distinctive scales, $\mu_H$ and $\mu_J$. The hard functions $H_{ab}^c$ live at the scale of the hard-scattering event,
\bea
\label{eq:muH}
\mu_H\sim p_T \,,
\eea
while the characteristic momentum scale for the siJMFs ${\cal G}_c$ is set by the jet dynamics and it is given by
\bea
\label{eq:muJ}
\mu_J \sim p_T R\,.
\eea
When $R\ll 1$, the dynamics at these two distinctive scales will not interfere with each other and thus factorize. This is the intuitive argument behind the factorization formalism in Eq.~\eqref{eq:fac-ungroom}. This first step of the factorization, the hard collinear factorization, is illustrated on the left-hand side of Fig.~\ref{fig:fac-ungroom}. We note that even though the factorized form of the cross section is derived for strictly $R\ll 1$, it was found that the factorization is a very good approximation even for fat jets with a relatively large jet radius of $R\sim 0.7$ and even above~\cite{Jager:2004jh,Mukherjee:2012uz,Dasgupta:2016bnd,Liu:2018ktv}, as pointed out also in \cite{Kolodrubetz:2016dzb}. These observations imply that the power corrections of the form ${\mathcal O}(R^2)$ to the factorization theorem in Eq.~(\ref{eq:fac-ungroom}) are in fact very small. While we do not have a general theoretical argument on the size of the power corrections,  we further verify in Sec.~\ref{sec:numerics} below that numerically this is indeed the case.
 
\bef
\includegraphics[height=2.8in]{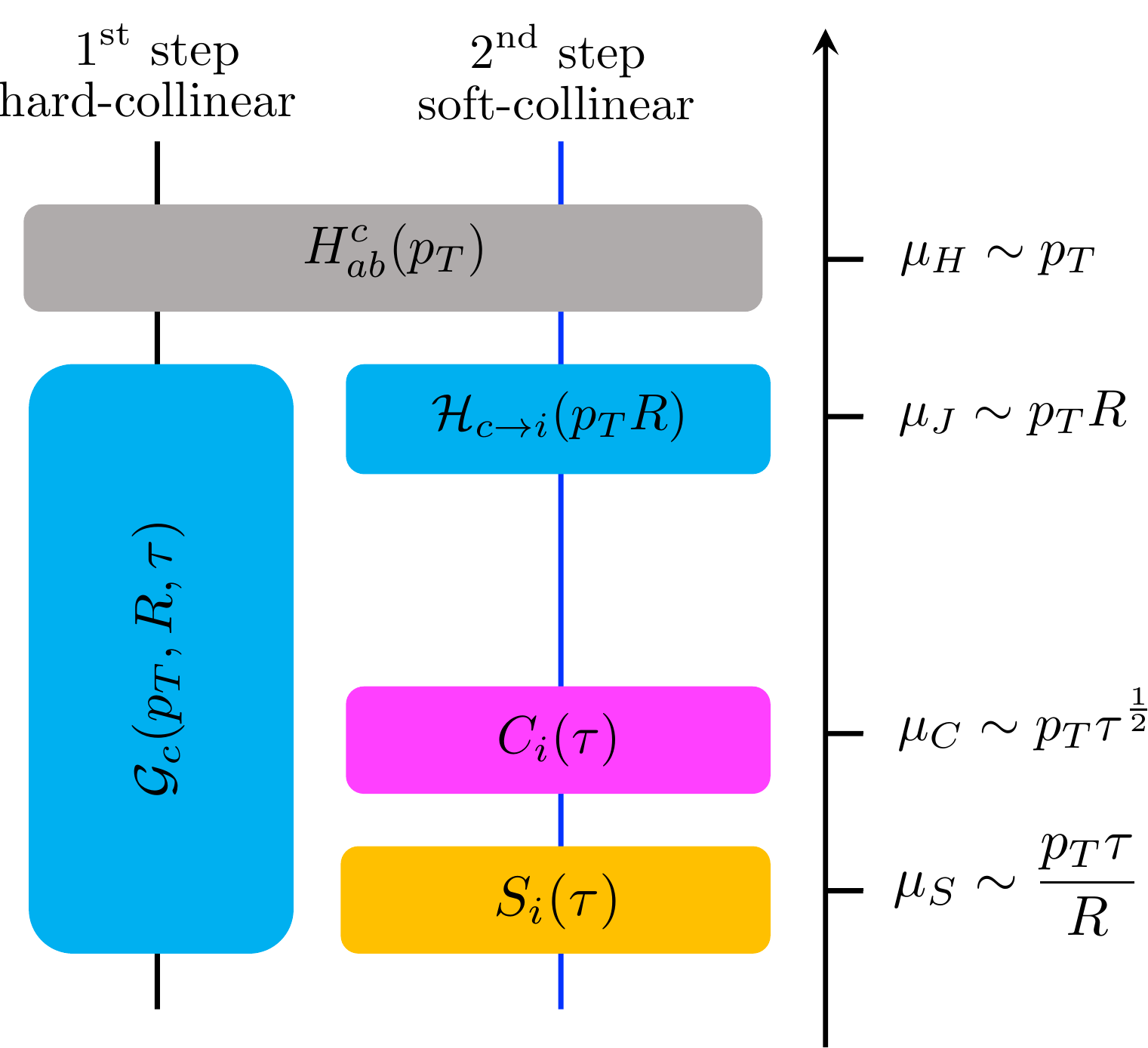} 
\caption{Illustration of the two-step factorization procedure for the ungroomed jet mass distribution. The first step is a hard collinear factorization of the siJMFs ${\cal G}_c$ from the hard functions $H_{ab}^c$. The second step is a soft collinear factorization of ${\cal G}_c$ in terms of hard matching functions ${\cal H}_{c\to i}$, collinear functions $C_i$, and collinear-soft functions $S_i$. \label{fig:fac-ungroom}}
\eef

We find that the siJMFs ${\cal G}_c$, as well as the corresponding hard functions $H_{ab}^c$, follow the usual timelike DGLAP evolution equations, which is consistent with the hard collinear factorization. See also~\cite{Kang:2016mcy,Dasgupta:2016bnd,Dai:2016hzf}. We find
\bea
\label{eq:DGLAP-G-ungroom}
\mu \f{d}{d\mu}{\cal G}_i(z,p_T, R,\tau,\mu)  = \sum_j \int_z^1 \f{dz'}{z'}P_{ji}\left(\f{z}{z'},\mu\right) {\cal G}_j(z',p_T, R,\tau,\mu)  \,,
\eea
where the $P_{ji}(z, \mu)$ denote the usual Altarelli-Parisi splitting functions. The DGLAP evolution equations for the siJMFs ${\cal G}_c$ enable us to resum single logarithms in the jet size parameter $\alpha_s^n\ln^{k} R$ with $k\leq n$ which is achieved by evolving the siJMFs from the jet scale $\mu_J\sim p_T R$ to the hard scale $\mu_H\sim p_T$~\cite{Vogt:2004ns,Anderle:2015lqa}. 

\subsection{Second step: soft collinear factorization}
The factorization formalism in Eq.~(\ref{eq:fac-ungroom}) is only valid for $\tau \sim R^2$. When $\tau$ is parametrically much smaller than the jet radius squared $\tau\ll R^2$,  the jet mass distribution receives additional large logarithmic corrections originating from soft and collinear emissions that need to be resummed to all orders.
In the small $\tau$ region, a second step of the factorization is required to resum logarithms of the form $\alpha_s^n \ln^{2n-k}\left(\tau/R^2\right)$ with $0\leq k \leq 2n $. This can be achieved by introducing two additional modes which follow the jet mass constraints: collinear modes within the jet and the collinear-soft mode~\cite{Bauer:2011uc,Chien:2015cka}. The collinear radiation within the jet has the momentum scaling 
\bea
p_c = (p_c^-, p_c^+, p_{c\perp}) \sim p_T(1, \lambda^2, \lambda)\,,
\label{eq:pc}
\eea
where $\lambda\sim \tau^{1/2}$. The collinear-soft radiation indicated by the subscript ``cs'' has the following momentum scaling
\bea
\label{eq:ps}
p_{cs}= (p_{cs}^-, p_{cs}^+, p_{cs\perp}) \sim \frac{p_T\tau}{R^2}\left(1, R^2, R\right).
\eea
At the same time, any hard-collinear emission of the order $p_T R$ has to be outside the jet as they would otherwise violate the hierarchy $\tau \ll R^2$, and thus do not contribute to the jet mass. In summary we obtain the following factorization for the siJMFs~\cite{Kang:2018qra}
\bea
\label{eq:refactorize}
{\cal G}_c(z,p_T, R,\tau,\mu) =& \sum_i \mathcal{H}_{c\to i}(z,p_T R,\mu) 
\nnu
&\hspace*{-2cm}\times\int d\tau^{C_i}d\tau^{S_i}\, \delta(\tau - \tau^{C_i}- \tau^{S_i})\, C_i\left(\tau^{C_i},p_T,\mu\right)
S_i\left(\tau^{S_i},p_T,R,\mu\right) \,.
\eea
Here $\mathcal{H}_{c\to i}(z, p_T R, \mu)$ are the hard matching functions, and describe how an energetic parton $c$ coming from the hard-scattering event produces a jet initiated by parton $i$ with radius $R$ carrying an energy fraction $z$ of the initial parton $c$. They are related to unconstrained radiation outside the jet, and thus they have the characteristic momentum scale $\mu_J\sim p_T R$ which is the jet scale. The relevant perturbative expressions and their renormalization group (RG) equations can be found in~\cite{Kang:2017mda,Kang:2018qra}. The collinear functions $C_i\left(\tau,p_T,\mu\right)$ that take into account collinear radiation inside the observed jet have the following definition at the operator level
\bea
C_{q}(\tau, p_T, \mu) =& \frac{1}{2N_c}
{\rm Tr} 
\bigg[\frac{\sla{\bar n}}{2}
\langle 0| \delta\left(\tau - \hat \tau^n \right)  \chi_n(0)  |JX\rangle
\langle JX|\bar \chi_n(0) |0\rangle \bigg]\,,
\\
C_{g}(\tau, p_T, \mu) =& - \frac{\omega}{2(N_c^2-1)}
\langle 0|  \delta\left(\tau - \hat \tau^{n} \right) {\mathcal B}_{n\perp \mu}(0) 
 |JX\rangle \langle JX|{\mathcal B}_{n\perp}^\mu(0)  |0\rangle\,.
\eea
The operator $\hat \tau^n$ is defined to count only the collinear radiation inside the jet. In fact up to an overall normalization, the collinear functions are the same as the usual inclusive jet functions, which describe the measurement of the invariant mass of the jet~\cite{Bauer:2001yt,Ritzmann:2014mka}. The corresponding perturbative results are available at next-to-leading order (NLO)~\cite{Bauer:2003pi,Fleming:2003gt} and next-to-next-to-leading order (NNLO)~\cite{Becher:2006qw,Becher:2010pd}. For completeness, we list here the results for renormalized collinear functions for quarks and gluons $i=q, g$ at NLO
\bea
C_i(\tau,p_T,\mu) &= \left(1+\f{\alpha_s}{2\pi}f_i \right)\delta(\tau)-\f{\alpha_s}{2\pi}\Bigg[\gamma_i \f{p_T^2}{\mu^2}\left(\f{\mu^2}{p_T^2 \tau}\right)_+ - 2C_i \f{p_T^2}{\mu^2} \left(\f{\mu^2}{p_T^2\tau} \ln\left(\f{p_T^2\tau}{\mu^2}\right)\right)_+\Bigg]\,.
\eea
Here we adopted the notation $C_i = C_{F,A}$ for quarks and gluons, respectively. The relevant constants $f_i$ and $\gamma_i$ are given by
\bea
f_q &= C_F\Bigg(\f{7}{2}-\f{\pi^2}{2}\Bigg)\, ,
\qquad
\gamma_q = \f{3C_F}{2}\,, 
\\
f_g &= C_A\Bigg(\f{67}{18}-\f{\pi^2}{2}\Bigg) -T_R N_f \f{10}{9}\, ,
\qquad
\gamma_g = \f{\beta_0}{2} \,.
\eea
From the perturbative NLO results one finds that the characteristic scale of the collinear functions is given by the jet mass which eliminates all large logarithms at a fixed order
\bea
\label{eq:muC}
\mu_C\sim p_T \tau^{\f{1}{2}} = m_J \,.
\eea
The collinear functions satisfy the following RG equations
\bea
\label{eq:collinearRG}
\mu \f{d}{d\mu} C_i(\tau, p_T,\mu) = \int d\tau'\, \gamma_{C_i}(\tau - \tau',p_T,\mu)\, C_i(\tau',p_T,\mu) \,,
\eea
where the anomalous dimensions $\gamma_{C_i}(\tau,p_T,\mu)$ are given by
\bea
\gamma_{C_i}(\tau,p_T,\mu) = \f{\alpha_s}{\pi}\left[\left( 2 C_i \ln\f{\mu^2}{p_T^2}+\gamma_i\right)\delta(\tau) - 2C_i\left(\f{1}{\tau}\right)_+\right]\,.
\eea
The soft functions $S_i(\tau, p_T, R, \mu)$ that appear in the factorized expression of the siJMFs in Eq.~(\ref{eq:refactorize}) have the following operator definitions
\bea
S_q(\tau, p_T, R, \mu ) =& \frac{1}{N_c} \langle 0| {\bar Y}_n  \, \delta(\tau - \hat \tau^s)
Y_{\bar{n} }  | X\rangle \langle  X|{\bar Y}_{\bar n}  Y_n  |0\rangle \,,
\\
S_g(\tau, p_T, R, \mu ) =& \frac{1}{N_c^2 - 1} \langle 0| {\bar {\cal Y}}_n  \, \delta(\tau - \hat \tau^s)
{\cal Y}_{\bar{n} }  | X\rangle \langle  X|{\bar {\cal Y}}_{\bar n}  {\cal Y}_n  |0\rangle\,.
\eea
Here $Y_n$ (${\cal Y}_n$) is a soft Wilson line in the fundamental (adjoint) representation along the light-like direction $n^\mu$ of the jet, while $Y_{\bar n}$ (${\cal Y}_{\bar n}$) is along the conjugated direction $\bar n^\mu$. Note that the operator $\hat \tau^s$ is defined to count only the soft radiation following the momentum scaling determined in Eq.~(\ref{eq:ps}). The perturbative results for the renormalized soft functions at NLO are given by
\bea
\label{eq:ren_S_ungroom}
S_i(\tau, p_T, R, \mu) &= \delta(\tau) + \f{\alpha_s C_i}{\pi}\Bigg[\f{\pi^2}{24}\delta(\tau) - 2\f{p_T}{\mu R}\left(\f{\mu R}{p_T \tau}\ln\left(\f{p_T \tau}{\mu R}\right)\right)_+\Bigg] \,,
\eea
from which the natural momentum scale is obtained to be
\bea
\label{eq:soft-scale}
\mu_S&\sim \f{p_T \tau}{R} = \frac{m_J^2}{p_T R} \,.
\eea
The corresponding RG equations are given by
\bea
\mu \f{d}{d\mu} S_i(\tau,p_T,R,\mu) = \int d\tau' \; \gamma_{S_i}(\tau - \tau',p_T,R,\mu) S_i(\tau',p_T,R,\mu)\,,
\eea
where the anomalous dimensions $\gamma_{S_i}(\tau,p_T,R, \mu)$ are given by
\bea
\label{eq:gamma_S_ungroom}
\gamma_{S_i}(\tau,p_T,R, \mu) = \f{\alpha_s C_i}{\pi}\left[2\left(\f{1}{\tau}\right)_+-\ln\left(\f{\mu^2 R^{2}}{p_T^2}\right)\delta(\tau) \right] \,.
\eea

\section{Factorization: the groomed jet mass \label{sec:groom-fac}}
In this section, we derive the factorization formalism for the soft drop groomed jet mass distribution for the single inclusive jet production in $pp$ collisions. We first give a brief review on the soft drop grooming algorithm, and then derive the corresponding factorized expression that allows for the resummation of all relevant large logarithmic corrections.

\subsection{Soft drop grooming \label{sec:softdrop}}
The soft drop grooming procedure recursively removes soft wide-angle radiation from an identified jet~\cite{Larkoski:2014wba}. The algorithm starts by re-clustering the constituents of an anti-$k_T$ jet~\cite{Cacciari:2008gp} with Cambridge-Aachen algorithm~\cite{Dokshitzer:1997in,Wobisch:1998wt} which yields an angular ordered clustering tree. One then steps backward through the clustering history of the jet and one iteratively removes soft branches from the jet. At each de-clustering step the jet is separated into two subjets or branches (also referred as proto-jets) with an angular separation $\Delta R_{ij} = \sqrt{(\Delta \eta)^2 +(\Delta \phi)^2}$ in the $\eta$-$\phi$ plane and transverse momenta $p_{Ti,j}$. At each step the following soft drop grooming criterion is checked
\bea
\label{eq:SD}
\frac{{\rm min}\left[p_{Ti}, p_{Tj}\right]}{p_{Ti}+p_{Tj}} > z_{\rm cut}\left(\frac{\Delta R_{ij}}{R}\right)^\beta \,.
\eea
The soft drop algorithm depends on two parameters: a soft threshold $z_{\rm cut}$ and an angular exponent $\beta$. Here $z_{\rm cut}$ sets the energy scale below which soft branches are removed from the jet. A typical value currently used by the experiments is $z_{\rm cut} = 0.1$. The parameter $\beta$ determines the sensitivity of the grooming algorithm to the wide-angle soft radiation. If the above criterion is not satisfied, the branch with the smaller $p_T$ is removed from  the jet. The procedure continues until the soft drop criterion is satisfied. The mass of the resulting groomed jet is usually referred to as the soft drop groomed jet mass which we denote by $m_{J,\rm gr}$. Correspondingly, we define the soft drop groomed $\tau_{\rm gr}$ measurement as
\bea\label{eq:taugroomed}
\tau_{\rm gr} = \frac{m_{J,\rm gr}^2}{p_T^2}\,. 
\eea
Note that in the denominator we still use the ungroomed jet transverse momentum $p_T$, instead of the $p_T$ of the groomed jet. This is because the ungroomed jet $p_T$ is an infrared and collinear (IRC) safe quantity, whereas the groomed analog is not IRC safe. See for example~\cite{Marzani:2017mva}. For $\beta=0$ the soft drop grooming algorithm corresponds to the modified mass drop tagger (mMDT)~\cite{Dasgupta:2013ihk}. Taking the limit $\beta\to \infty$ removes the groomer and the ungroomed jet mass distribution is recovered. We are going to discuss this limit in more detail below.

\subsection{First step: hard collinear factorization}
Following our discussion of the ungroomed case, the first step factorization for the groomed jet mass distribution takes the form
\bea
\label{eq:fac-groom}
\f{d\sigma}{d\eta dp_T d\tau_{\rm gr}} =& \sum_{abc} f_a(x_a,\mu) \otimes f_b(x_b,\mu) \otimes H_{ab}^c(x_a,x_b,\eta,p_T/z,\mu)
\nnu 
&\otimes \,{\cal G}_c^{\rm gr}(z,p_T,R, \tau_{\rm gr},\mu; z_{\rm cut}, \beta)\;.
\eea
Here ${\cal G}_c^{\rm gr}$ are the groomed siJMFs that take into account the soft drop groomed jet mass measurement $\tau_{\rm gr}$ for the observed jet. The groomed siJMFs have the following slightly modified operator definitions
\bea
{\cal G}_q^{\rm gr}(z,p_T, R,\tau_{\rm gr},\mu; z_{\rm cut}, \beta) =& \f{z}{2N_c}{\rm Tr} \big[\f{\sla{\bar n}}{2}
\langle 0| \delta\left(\omega - \bar n\cdot {\mathcal P} \right)\delta(\tau_{\rm gr} - \hat \tau_{\rm gr}(J)) \chi_n(0)  
\nnu
&\times
|JX\rangle \langle JX|\bar \chi_n(0) |0\rangle \big]\,,
\\
{\cal G}_g^{\rm gr}(z,p_T, R,\tau_{\rm gr},\mu; z_{\rm cut}, \beta) =& - \f{z\,\omega}{2(N_c^2-1)}
\langle 0|  \delta\left(\omega - \bar n\cdot {\mathcal P} \right)\delta(\tau_{\rm gr} - \hat \tau_{\rm gr}(J)) {\mathcal B}_{n\perp \mu}(0) 
\nnu
&\times
 |JX\rangle \langle JX|{\mathcal B}_{n\perp}^\mu(0)  |0\rangle\,,
\eea
where the operator $\hat \tau_{\rm gr}(J)$ represents the jet mass measurement in the presence of soft drop grooming as described above, with the measured value being equal to $\tau_{\rm gr}$. This first step of the factorization in Eq.~\eqref{eq:fac-groom} is conceptually the same as Eq.~\eqref{eq:fac-ungroom} for the ungroomed case, where only the ungroomed siJMFs ${\cal G}_c(z,p_T,R, \tau,\mu)$ are replaced by their corresponding groomed analog ${\cal G}_c^{\rm gr}(z,p_T,R, \tau_{\rm gr},\mu; z_{\rm cut}, \beta)$. This factorization holds in the region in which $z_{\rm cut} \sim 1 $ and $\tau_{\rm gr} \sim R^2$. 

We note that the standard jet transverse momentum $p_T$ is set by the hard scattering dynamics at this step, i.e. associated with the hard functions $H_{ab}^c$ in the above factorization theorem, which is the same as that for the ungroomed jet mass factorization. Therefore it is consistent to use the ungroomed jet $p_T$ also for the case of the groomed jet mass distribution in Eq.~(\ref{eq:taugroomed}). From the universality of the factorization formalism, the RG equations for the groomed siJMFs have to be consistent with that of the hard functions $H_{ab}^c$, and thus are the same as for the ungroomed case. Therefore, the groomed siJMFs satisfy again the usual DGLAP evolution equations that can be used to resum single logarithms in the jet size parameter
\bea
\label{eq:DGLAP-G-groom}
\mu \f{d}{d\mu}{\cal G}_i^{\rm gr}(z,p_T, R,\tau_{\rm gr},\mu; z_{\rm cut}, \beta)  = \sum_j \int_z^1 \f{dz'}{z'}P_{ji}\left(\f{z}{z'},\mu\right) {\cal G}_j^{\rm gr}(z',p_T, R,\tau_{\rm gr},\mu; z_{\rm cut}, \beta)  \,. 
\eea

\subsection{Second step: soft collinear factorization with soft drop grooming \label{sec:scfacgr}}
In practice, the LHC experiments usually choose $z_{\rm cut} \sim 0.1$ while $\tau_{\rm gr}$ can be as low as ${\cal O}\left( 10^{-4} \right)$. A typical value for the jet radius parameter is $R = 0.8$, see section~\ref{sec:numerics} below.
Therefore, we are particularly interested in the factorization of the cross section in the region where $\tau_{\rm gr} / R^2 \ll z_{\rm cut} \ll 1$. In~\cite{Marzani:2017mva} finite $z_{\rm cut}$ corrections were considered which turn out to be very small for all practical purposes. Following a similar discussion as in~\cite{Frye:2016aiz}, we now focus on the refactorization of the siJMFs in the presence of soft drop grooming. We start by identifying the relevant modes in order to derive a factorization theorem in the kinematic region of interest. Similar to the ungroomed case we have $\tau_{\rm gr}/R^2 \ll 1$ which implies that only collinear and soft degrees of freedom are relevant to leading power. Therefore, in order to closely relate our discussion here with the ungroomed jet mass distribution discussed above, we study in detail how the soft drop grooming algorithm modifies the factorized structure obtained in Eq.~\eqref{eq:refactorize}.

Any hard collinear radiation at the scale $\mu_J\sim p_T R$ is captured by the hard matching functions ${\cal H}_{c\to i}(z, p_T R, \mu)$. They correspond to energetic out-of-jet radiation contributions which are not affected by the soft drop grooming algorithm that only deals with the in-jet dynamics. Therefore, the hard matching functions ${\cal H}_{c\to i}$ are not modified in the presence of soft drop grooming.

The collinear radiation inside the jet is described by the collinear functions $C_i(\tau, p_T, \mu)$. To leading power in $z_{\rm cut}$, the collinear functions are also not modified by the soft drop grooming algorithm as all the energetic collinear in-jet radiation always passes the soft drop criterion. This can be understood as follows. Let us denote the energy fraction of the softer branch after a de-clustering step by $z$. For collinear modes, $z$ should generally satisfy $z\sim 1 \gg z_{\rm cut}$ which means that the branch passes the soft drop criterion. The dynamics of other branches with the scaling $z\sim z_{\rm cut}\ll 1$ are naturally captured by the soft functions.
\bef
\includegraphics[height=2.3in,angle=-45,]{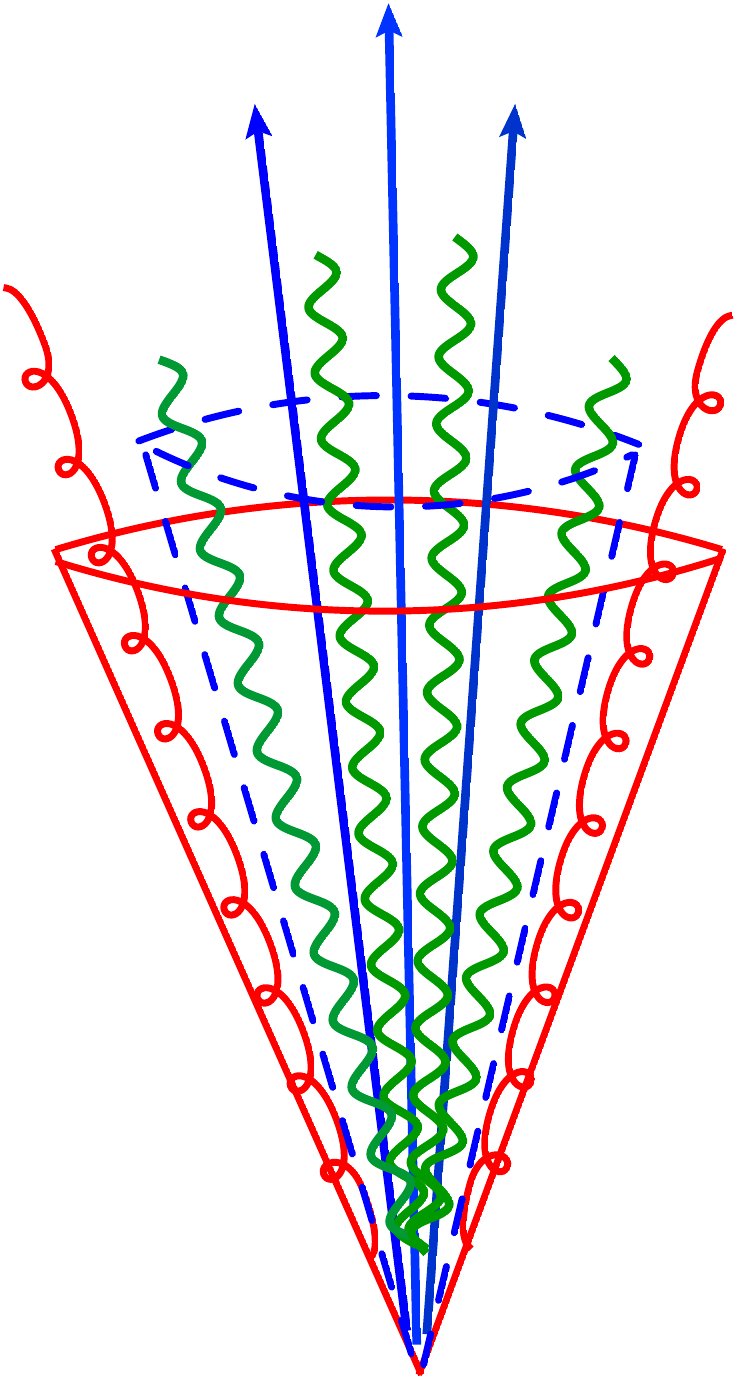} 
\caption{Schematic illustration of the relevant modes for the refactorized siJMFs when soft drop grooming is included. The green lines represent the soft mode that passes the soft drop criterion. Soft radiation at larger angles that fails the soft drop criterion is illustrated by the red lines. The blue lines represent the more energetic collinear radiation inside the jet which is not affected by the soft drop grooming algorithm up to power corrections. The dashed blue jet cone represents the groomed jet obtained from the ungroomed jet shown in red. \label{fig:groomed_jet}}
\eef
In the analytical calculations, one can show that after zero-bin subtraction~\cite{Manohar:2006nz}, the $z_{\rm cut}$-dependent contributions to the collinear functions are suppressed by powers of $z_{\rm cut}$. Since we are working in the parametric limit $z_{\rm cut} \ll 1$, one may safely neglect these power corrections of order ${\cal O}(z_{\rm cut})$. The situation here is in fact very similar to the jet angularity calculations in~\cite{Ellis:2010rwa}. One finds that the jet algorithm leads to a constraint on the parton branching fraction $z$ such that $z_{\rm lim} < z< 1-z_{\rm lim}$, with $z_{\rm lim}\sim \tau_a/R^2$. As it was shown carefully in~\cite{Ellis:2010rwa}, these constraints lead to $z_{\rm lim}$-dependent contributions, which are power suppressed precisely by $z_{\rm lim}$ when $\tau_a/R^2 \ll 1$. The role of $z_{\rm lim}$ in the angularity calculation is now replaced by $z_{\rm cut}$~\cite{Marzani:2017mva} from the soft drop grooming algorithm, and thus the same conclusions hold.

Finally, let us consider the soft radiation. We find that the soft radiation (or collinear-soft mode) contains particles which may or may not pass the soft drop grooming criterion. Since we are working with the hierarchy $\tau_{\rm gr}/R^2 \ll z_{\rm cut}$, soft radiation emitted at a relatively large angle will naturally fail the soft drop criterion. This can be understood as follows. We choose to work in a reference frame where the jet has no transverse momentum component and let us denote its large light-cone component by $\omega_J$. Now we consider the situation where the soft particle with momentum $k$ and $z = k^0/E_J = \left(k^+ + k^-\right)/\omega_J$ is radiated at an angle $\theta$ with respect to the jet axis. The soft drop criterion can then be written as
\bea
\label{eq:criterion}
z > z_{\rm cut} \left(\frac{\theta}{R}\right)^\beta\,. 
\eea
For the large angle soft radiation inside the jet, we have  
\bea
\label{eq:algorithm}
\frac{k^+}{k^-}\sim \theta^2 \lesssim R^2\,.
\eea
If the soft radiation passes the soft drop criterion in Eq.~\eqref{eq:criterion}, they would remain in the final groomed jet, and thus contribute to the jet mass observable, 
\bea
\label{eq:mass}
m_{J,\rm gr}^2 \sim \omega_J k^+\,.
\eea
Combining the above Eqs.~\eqref{eq:criterion},~\eqref{eq:algorithm}, and \eqref{eq:mass}, one would have
\bea
\tau_{\rm gr} / R^2 \gtrsim z_{\rm cut}\,, 
\eea
which violates the hierarchy $\tau_{\rm gr} / R^2 \ll z_{\rm cut}$. Therefore, the soft radiation at relatively large angles inside the jet will not contribute to the groomed jet mass $m_{J, \rm gr}$ or $\tau_{\rm gr}$. The precise momentum scaling of these soft emissions is given by
\bea
p_{cs}^{\notin{\rm gr}} \sim z_{\rm cut} p_T\left(1, R^2, R\right)\,,
\eea
where the superscript ``$\notin\!\!\!{\rm gr}$'' emphasizes the fact that they do not pass the soft drop grooming criterion. 

\bef
\includegraphics[height=2.8in]{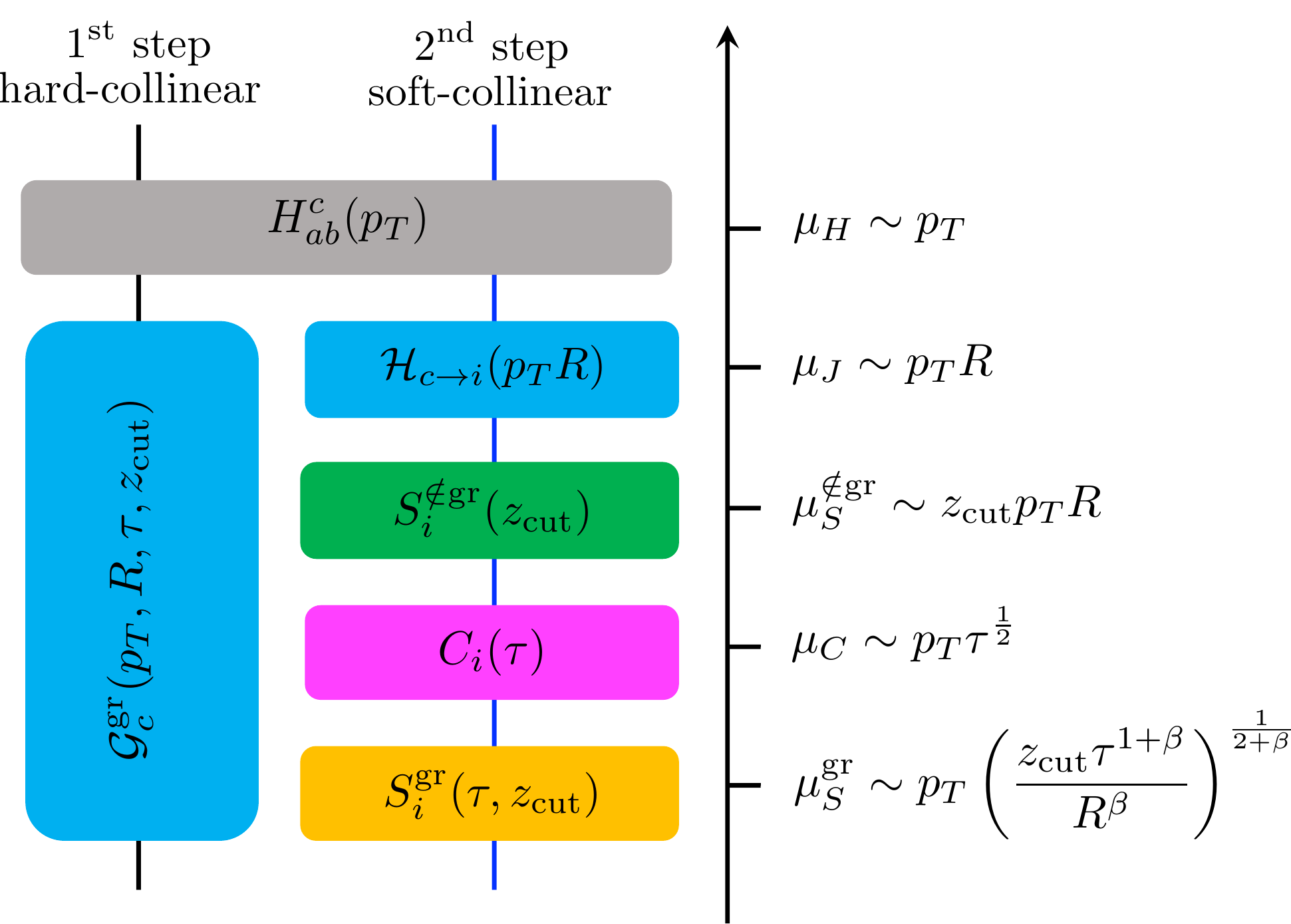} 
\caption{Illustration of the two-step factorization procedure for the jet mass distribution in the presence of soft drop grooming. The first step is a hard collinear factorization of the groomed siJMFs ${\cal G}_c^{\rm gr}$ from the hard functions $H_{ab}^c$. The second step is a soft collinear factorization of ${\cal G}_c^{\rm gr}$ in terms of hard matching functions ${\cal H}_{c\to i}$, collinear functions $C_i$ as well as two soft functions $S_i^{\notin{\rm gr}}$ and $S_i^{\rm gr}$. Here $S_i^{\notin{\rm gr}}$ captures soft emissions at relatively large angles within the jet which fail the soft drop criterion, whereas $S_i^{\rm gr}$ contains collinear soft radiation that passes the soft drop criterion. \label{fig:fac-groom}}
\eef

On the other hand, the soft radiation that is emitted at smaller angles $\theta \ll R$ passes the soft drop criterion and will contribute to the observed groomed jet mass $\tau_{\rm gr}$. In this case, with $k^+/k^- \sim \theta^2 \ll R^2$, following the same analysis as above, we obtain the momentum scaling for the more collimated soft radiation
\bea
p_{cs}^{\rm gr} \sim \left(z_{\rm cut}\right)^{\frac{2}{2+\beta}}
\left(\frac{\tau}{R^2}\right)^{\frac{\beta}{2+\beta}} p_T
\left(1, \left(\frac{\tau R^\beta}{z_{\rm cut}}\right)^{\frac{2}{2+\beta}}, \left(\frac{\tau R^\beta}{z_{\rm cut}}\right)^{\frac{1}{2+\beta}}\right)\,.
\eea
Here the superscript ``${\rm gr}$'' emphasizes the fact that the soft radiation passes the soft drop criterion and will thus end up within the final groomed jet. To summarize, we have the following refactorized expression for the groomed siJMFs
\bea
\label{eq:refactorize-groom}
{\cal G}_c^{\rm gr}(z,p_T, R, \tau_{\rm gr}, \mu; z_{\rm cut}, \beta) =& \sum_i \mathcal{H}_{c\to i}(z, p_T R, \mu) S_i^{\notin{\rm gr}} (p_T, R, \mu; z_{\rm cut}, \beta)
\nnu
&\hspace{-35mm}
\times  \int d\tau^{C_i}d\tau^{S_i} \delta(\tau_{\rm gr}- \tau^{C_i}- \tau^{S_i}) \,C_i(\tau^{C_i}, p_T, \mu)\,S_i^{\rm gr}(\tau^{S_i}, p_T, R, \mu; z_{\rm cut}, \beta)\,.
\eea
Here the hard matching functions $\mathcal{H}_{c\to i}$ and the collinear functions $C_i(\tau, p_T, \mu)$ are the same as for the ungroomed case, see Eq.~(\ref{eq:refactorize}) above. However, the soft functions are different where $S_i^{\notin{\rm gr}}$ takes into account soft radiation that fails the soft drop criterion, while $S_i^{\rm gr}$ is associated with soft particles that pass the soft drop criterion and thus remain inside the groomed jet. We note that the $R$ dependence of the soft function $S_i^{\rm gr}$ is only due to the soft drop constraint in Eq.~(\ref{eq:SD}) instead of the jet clustering constraint. We further illustrate the factorization in Eqs.~\eqref{eq:fac-groom} and \eqref{eq:refactorize-groom} in Figs.~\ref{fig:groomed_jet} and \ref{fig:fac-groom}. 

When we consider the kinematic region with the scaling $\tau_{\rm gr}/R^2\sim z_{\rm cut}\ll 1$, there is a transition point~\cite{Larkoski:2014wba,Baron:2018nfz} at $\tau_{\rm gr}/R^2=z_{\rm cut}$ above which the groomed factorization theorem in Eq.~(\ref{eq:refactorize-groom}) is reduced to the ungroomed case, as outlined in section~\ref{sec:groomed_jet}. We present the detailed derivation in the Appendix~\ref{app:softpass}. The numerical results also show the existence of the transition point between the groomed and the ungroomed case at large values of the jet mass as presented in section~\ref{sec:groomed_jet}.

\subsection{Soft functions at NLO}
In this section, we present the explicit NLO expressions for both types of soft functions that appear in the factorization theorem in Eq.~(\ref{eq:refactorize-groom}). We refer the interested reader to the Appendix for a more detailed derivation. The soft functions $S_i^{\notin{\rm gr}}$ do not depend on the groomed jet mass $\tau_{\rm gr}$. The reason is that they only take into account soft particles that fail the soft drop criterion and, hence, those soft particles do not contribute to the groomed jet mass. Up to NLO, the renormalized soft functions $S_i^{\notin{\rm gr}}$ for quarks and gluons $i=q,g$ are given by
\bea
\label{eq:ren_S_notgr}
S_i^{\notin{\rm gr}}(p_T, R, \mu; z_{\rm cut}, \beta) = 1+ \frac{\alpha_s}{2\pi} \frac{C_i}{1+\beta} \left[\frac{1}{2}\ln^2\left(\frac{\mu^2}{z_{\rm cut}^2 p_T^2 R^2}\right) - \frac{\pi^2}{12}\right],
\eea
which is independent of $\tau_{\rm gr}$ as expected. From the above result, one can obtain the natural momentum scale for $S_i^{\notin{\rm gr}}$, which is given by
\bea
\label{eq:notingroom}
\mu_{S}^{\notin{\rm gr}} = z_{\rm cut} p_T R\,.
\eea
The RG equations of $S_i^{\notin{\rm gr}}$ are multiplicative and take the following form
\bea
\mu \f{d}{d\mu} S_i^{\notin{\rm gr}} (p_T, R, \mu; z_{\rm cut}, \beta) = \gamma_{S_i}^{\notin{\rm gr}}(p_T, R, \mu; z_{\rm cut}, \beta) S_i^{\notin{\rm gr}} (p_T, R, \mu; z_{\rm cut}, \beta)\,.
\eea
The relevant anomalous dimensions $\gamma_{S_i}^{\notin{\rm gr}}$ are given by
\bea
\label{eq:gamma_S_notgr}
\gamma_{S_i}^{\notin{\rm gr}}(p_T, R, \mu; z_{\rm cut}, \beta)  = \frac{\alpha_s}{\pi} \frac{C_i}{1+\beta} \ln\left(\frac{\mu^2}{z_{\rm cut}^2 p_T^2 R^2}\right)\,.
\eea
The other soft functions $S_i^{\rm gr}$ describe the soft radiation that passes the soft drop criterion and therefore contributes to the groomed jet mass. The renormalized soft functions $S_i^{\rm gr}$ up to NLO are given by
\bea
\label{eq:ren_S_gr}
S_i^{\rm gr}(\tau, p_T, R, \mu; z_{\rm cut}, \beta) = \delta(\tau) + \frac{\alpha_s}{\pi} C_i \left[
\frac{\pi^2}{24}\frac{2+\beta}{1+\beta} \delta(\tau) - \frac{2(1+\beta)}{2+\beta} A \left(\frac{\ln\left(A\tau\right)}{A\tau} \right)_+ \right]\;,
\eea
where the factor $A$ is given by the following expression
\bea
A =\left[ \left(\frac{z_{\rm cut}}{R^\beta}\right)^{\frac{1}{2+\beta}} \frac{p_T}{\mu}\right]^{\frac{2+\beta}{1+\beta}}\,.
\eea
From the perturbative NLO result, we find that the natural scale for $S_i^{\rm gr}$ is
\bea
\label{eq:soft-scale-groom}
\mu_S^{\rm gr} = p_T \left( \frac{z_{\rm cut} \tau^{1+\beta}}{R^\beta}\right)^{\frac{1}{2+\beta}}\,.
\eea
The associated RG equations have a convolution structure with respect to $\tau$ and are given by
\bea
\mu \f{d}{d\mu} S_i^{\rm gr}(\tau, p_T, R, \mu; z_{\rm cut}, \beta) =& \int d\tau' \; \gamma_{S_i}^{\rm gr}(\tau - \tau',p_T,R,\mu; z_{\rm cut}, \beta) 
\nnu
&\times S_i^{\rm gr}(\tau', p_T, R, \mu; z_{\rm cut}, \beta)\,,
\eea
where the anomalous dimensions $\gamma_{S_i}^{\rm gr}$ are given by
\bea
\label{eq:gamma_S_gr}
\gamma_{S_i}^{\rm gr}(\tau,p_T,R, \mu; z_{\rm cut}, \beta) = \f{2\alpha_s C_i}{\pi}\left[\left(\f{1}{\tau}\right)_++ \ln(A)
\delta(\tau) \right] \,.
\eea

\subsection{Consistency between the groomed and ungroomed case \label{sec:consistency}}

We are now going to study the connection between the factorization formalism for the groomed and ungroomed jet mass distribution which provides an important consistency check of the obtained factorization theorems. In the kinematic region of interest, $\tau/R^2 \ll z_{\rm cut}\ll 1$, we find that both the hard matching functions ${\cal H}_{c\to i}$ taking into account out-of-jet radiations and the collinear functions $C_i(\tau, p_T, \mu)$ are the same for both cases. As mentioned above, it turns out that only the soft functions are different. From the consistency of the RG evolution equations, one expects that the anomalous dimensions $\gamma_{S_i}^{\notin{\rm gr}}$ and~$\gamma_{S_i}^{\rm gr}$ for the groomed jet mass distribution should be related to the anomalous dimension $\gamma_{S_i}$ for the ungroomed case. In fact, the consistency between the two cases requires the soft anomalous dimensions to satisfy the following relation
\bea
\gamma_{S_i}^{\notin{\rm gr}}(p_T, R, \mu; z_{\rm cut}, \beta)\, \delta(\tau) + \gamma_{S_i}^{\rm gr}(\tau,p_T,R, \mu; z_{\rm cut}, \beta) = \gamma_{S_i}(\tau,p_T,R, \mu)\,.
\eea
From the explicit expressions given in Eqs.~\eqref{eq:gamma_S_notgr}, \eqref{eq:gamma_S_gr} and~\eqref{eq:gamma_S_ungroom} above, we can directly verify that the above equality indeed holds true.

When we take the limit $\beta \to \infty$, the soft drop criterion is always satisfied and we get back to the ungroomed jet mass distribution. This limit can be studied directly at the level of the perturbative NLO expressions of the soft functions presented above. One observes that the renormalized soft functions $S_i^{\text{gr}}$ in Eq. (\ref{eq:ren_S_gr}), and their anomalous dimensions $\gamma_{S_i}^{\text{gr}}$ in Eq. (\ref{eq:gamma_S_gr}) reduce to their ungroomed counterparts in the limit $\beta\to\infty$.  The relevant NLO results for the ungroomed soft functions $S_i$ can be found in Eq.~(\ref{eq:ren_S_ungroom}), and their anomalous dimensions $\gamma_{S_i}$ in Eq. (\ref{eq:gamma_S_ungroom}). In addition, the renormalized soft functions $S_i^{\notin \text{gr}}$, Eq. (\ref{eq:ren_S_notgr}), and their anomalous dimensions $\gamma_{S_i}^{\notin{\rm gr}}$, Eq.~\eqref{eq:gamma_S_notgr}, approach $1$ and $0$, respectively, in the limit $\beta \to \infty$. In section~\ref{sec:groomed_jet}, we are also going to study the transition between the groomed and ungroomed jet mass distributions by taking the limit $\beta\to\infty$ numerically.

\subsection{Comment on non-global logarithms and comparison to the literature \label{sec:NGLs}}
Before presenting phenomenological results at the LHC, we would like to briefly comment on the role of NGLs for both the groomed and ungroomed jet mass distribution which we do not take into account in our factorization theorems above. In addition, we address in more detail how our new factorization formalism compares to results available in the literature. Generally, NGLs arise from gluons outside the jet that radiate soft gluons into the jet~\cite{Dasgupta:2001sh,Banfi:2002hw}. This leads to single logarithmic contributions starting at NNLO. In order to do precision jet substructure calculations such contributions have to be taken into account even though the NGL contribution is often found to be rather small. In the past years a lot of progress has been made in order to better understand the complicated all order structure of NGLs, see for example~\cite{Larkoski:2016zzc,Becher:2015hka,Larkoski:2015zka,Neill:2016stq,Becher:2016mmh,Caron-Huot:2015bja,Schwartz:2014wha}. The ungroomed jet mass distribution as discussed in section~\ref{sec:ungroom-fac} receives single logarithmic non-global contributions of the form  $\alpha_s^n \ln^k(\tau/R^2)$ with $k\leq n$. In this sense the NGLs directly affect the jet mass spectrum. For the groomed case, these logarithms of the jet mass are absent ($\beta=0$) or power suppressed ($\beta>0$). Note that for $\beta\to\infty$ the usual NGLs for the ungroomed case are reproduced. See~\cite{Larkoski:2014wba} for a more detailed discussion. However, also the groomed jet mass distribution receives corrections from NGLs which affect the absolute normalization of the cross section and also indirectly the groomed jet mass spectrum. For the groomed inclusive jet mass spectrum NGLs arise due to the angular correlation of emissions between the in-jet wide angle soft radiation in $S_i^{\notin \text{gr}}$ and the hard collinear radiation outside the jet in ${\cal H}_{c\to i}$\footnote{The logarithms due to the correlation between in- and out-of-jet radiation within the soft mode have already been captured by our factorization theorems.}. Therefore, there are NGLs of the form $\alpha_s^n\ln^k z_{\rm cut}$ with $k\leq n$ that will change the absolute normalization of the cross section. In addition, since NGLs affect the quark and gluon contributions differently they will also indirectly affect the shape of the groomed jet mass distribution. Of course, for all practical purposes the numerical effect of NGLs is expected to be rather small for both the groomed and ungroomed jet mass distribution unless $z_{\rm cut}$ is chosen to be very small.

Finally, we would like to compare our new approach to the jet mass distribution for inclusive jet production to results available in the literature. In particular, we compare to the results of~\cite{Frye:2016aiz}. See also~\cite{Marzani:2017kqd,Marzani:2017mva,Larkoski:2017cqq, Makris:2017arq} for example. In~\cite{Frye:2016aiz}, the inclusive groomed jet mass distribution was considered in $pp\to Z+\text{jet}+X$ events. The event topology considered in this work is therefore different $pp\to\text{jet}+X$ but the general factorization structure is the same. Using the notation developed in this work, the factorized structure employed in~\cite{Frye:2016aiz} can be summarized as follows
\be
\frac{d\sigma}{d\eta dp_T d\tau_{\rm gr}} = \sum_{c} H_c'(p_T,\eta,R,z_{\rm cut},\beta,\mu)\; C_c(\tau_{\rm gr}, p_T, \mu)\otimes S_c^{\rm gr}(\tau_{\rm gr}, p_T, R, \mu; z_{\rm cut}, \beta) \,,
\ee
where the sum is taken over $c=q,\bar q,g$. The hard functions defined here $H_c'$ are independent of $\tau_{\rm gr}$ and have been extracted in~\cite{Frye:2016aiz} to NLO from MCFM~\cite{Campbell:2002tg} for $pp\to Z+\text{jet}+X$. Instead, the factorization framework presented in this work now allows for a further separation of $H_c'$ in terms of hard functions $H_{ab}^c$, hard matching functions ${\cal H}_{c\to i}$ and soft functions $S_i^{\notin \text{gr}}$ taking into account soft radiation that fails the soft drop criterion, see Eqs.~(\ref{eq:fac-groom}) and~(\ref{eq:refactorize-groom}). The additional factorization allows for the resummation of logarithms in the jet size parameter $R$ and, more importantly, logarithms in the soft threshold parameter $z_{\rm cut}$ which otherwise can only be determined numerically to fixed order. An important feature of our new formalism is that by resumming all logarithms in $z_{\rm cut}$ we are able to reliably predict the absolute normalization of groomed jet observables, which was not achieved for $pp$ collisions before, up to NGLs. In addition, our new formalism in principle allows us to also systematically include NGLs for groomed jet observables since they can be clearly associated with certain parts of our factorization theorem as discussed above. However, numerical studies of NGLs are beyond the scope of this work and will be addressed in the future.

\section{Phenomenology at the LHC \label{sec:numerics}}
In this section, we present numerical results for jet mass distribution at LHC energies, for both ungroomed and soft drop groomed jets in $pp\to\text{jet}+X$. We first present details of our numerical studies and we then compare with the experimental data taken at the LHC. 

\subsection{RG evolution}
For all the numerical studies, we closely follow the methods used in the jet angularity paper of~\cite{Kang:2018qra}. We solve the respective evolution equations of the collinear and soft functions in position space for which we define the Fourier transform of a generic function $F$ depending on $\tau$ as
\bea\label{eq:Fourier}
F(x) = \int_0^\infty d\tau\, e^{-ix\tau} F(\tau)\,.
\eea
We then evolve the collinear and soft functions from their canonical scales to the jet scale $\mu_J\sim p_T R$ where they will be combined with the hard matching functions ${\cal H}_{c\to i}$ in order to obtain the siJMFs ${\cal G}_c$ in Eq.~\eqref{eq:refactorize} or their groomed counterparts ${\cal G}_c^{\rm gr}$ in Eq.~\eqref{eq:refactorize-groom}. For more details, see~\cite{Kang:2013lga, Kang:2018qra}. The final expressions for the ungroomed siJMFs ${\cal G}_c$ can be written in terms of the evolved collinear and soft functions as
\bea
\label{eq:Gevol-ungroom}
{\cal G}_c(z,p_T, R, \tau,\mu) =& \sum_i \mathcal{C}_{c\to i}(z,p_T R,\mu) \int \f{dx}{2\pi} e^{ix\tau} \exp\left[\int^{\mu_J}_{\mu_C}\frac{d\mu'}{\mu'} \gamma_{C_i}(x,p_T,\mu')\right]
\nnu
&\times\exp\left[\int^{\mu_J}_{\mu_S}\frac{d\mu'}{\mu'} \gamma_{S_i}(x,p_T,R,\mu')\right]C_i(x,p_T,\mu_C)S_i(x,p_T,R,\mu_S) \,,
\eea
where the convolution over $\tau$ becomes a simple product in the position space variable $x$. The coefficient functions $\mathcal{C}_{c\to i}(z,p_T R,\mu)$ are related to ${\cal H}_{c\to i}$ and their explicit expressions can be found in~\cite{Kang:2017glf}. The perturbative results of the relevant functions and their anomalous dimensions in position space can be derived by taking the Fourier transform following the definition in Eq.~\eqref{eq:Fourier}. It might be instructive to point out that the above RG running from $\mu_C\sim p_T\tau^{1/2}$ to $\mu_J$, as well as from $\mu_S\sim p_T\tau/R$ to $\mu_J$ are both resumming the logarithms of type~$\alpha_s^n\ln^{2k}\left(\tau/R^2\right)$ with $k\leq n$ at next-to-leading logarithmic (NLL) accuracy. Similarly, we can obtain the final expressions for the groomed siJMFs ${\cal G}_c^{\rm gr}$ in terms of the evolved functions in position space as
\bea
\label{eq:Gevol-groom}
{\cal G}_c^{\rm gr}(z,p_T, R, \tau_{\rm gr},\mu; z_{\rm cut}, \beta) =& \sum_i \mathcal{C}_{c\to i}(z,p_T R,\mu) 
\nnu
&\times
\exp\left[\int_{\mu_S^{\notin{\rm gr}}}^{\mu_J} \frac{d\mu'}{\mu'} \gamma_{S_i}^{\notin{\rm gr}}(p_T, R, \mu'; z_{\rm cut}, \beta) \right]
S_i^{\notin{\rm gr}}(p_T, R, \mu_S^{\rm gr}; z_{\rm cut}, \beta)
\nnu
&\times
 \int \f{dx}{2\pi} e^{ix\tau_{\rm gr}} \exp\left[\int^{\mu_J}_{\mu_C}\frac{d\mu'}{\mu'} \gamma_{C_i}(x,p_T,\mu')\right] C_i(x,p_T,\mu_C)
\nnu
&\times\exp\left[\int^{\mu_J}_{\mu_S^{\rm gr}}\frac{d\mu'}{\mu'} \gamma_{S_i}^{\rm gr}(x,p_T,R,\mu')\right]  S_i^{\rm gr}(x,p_T,R,\mu_S; z_{\rm cut}, \beta)\,.
\eea
Here, the RG evolution of the collinear function between the scales $\mu_C$ and $\mu_J$ resums logarithms in $\tau/R^2$ which is the same as in the ungroomed case. In addition, logarithms in $z_{\rm cut}$ that are introduced by the grooming procedure are resummed through the RG running as can be seen explicitly here. The soft function $S_i^{\notin{\rm gr}}$ is evolved from its characteristic scale from $\mu_S^{\notin{\rm gr}}\sim z_{\rm cut}p_T R$ to the jet scale $\mu_J\sim p_T R$ and similarly for $S_i^{{\rm gr}}$. The resummation of logarithms in $z_{\rm cut}$ is particularly important when $z_{\rm cut}$ is chosen to be very small. For our phenomenological results presented below we always choose $z_{\rm cut}=0.1$. See for example~\cite{Hoang:2017kmk} where the authors proposed to use $z_{\rm cut}$ values down to 0.001 which was termed ``light grooming''. The resummation of logarithms in $z_{\rm cut}$ is related to NGLs and is particularly relevant in order to determine the absolute normalization of the groomed jet cross section as discussed in more detail in section~\ref{sec:NGLs}.

With the above results for ${\cal G}_c$ and ${\cal G}_c^{\rm gr}$ at the canonical scale $\mu_J$, we further evolve the ungroomed/groomed siJMFs through their DGLAP equations in Eqs.~\eqref{eq:DGLAP-G-ungroom} and \eqref{eq:DGLAP-G-groom} from $\mu_J\sim p_T R$ to the hard scale $\mu_H\sim p_T$. This second step of the RG evolution resums single logarithms in the jet size parameter $R$. By solving all relevant RG evolution equations we are thus able to resum three dominant classes of logarithmic corrections to all orders in the strong coupling constant for the groomed jet mass: logarithms in the jet mass $\tau_{\rm gr}/R^2$, the jet radius $R$ and the soft threshold $z_{\rm cut}$. For the ungroomed case, there are no logarithms in $z_{\rm cut}$ and the jet mass logarithms are given in terms of $\tau/R^2$.

\subsection{Non-perturbative shape functions and scale variations \label{sec:shapefct}}
For small values of $\tau$, the soft scale $\mu_S\sim p_T\tau/R$ in Eq.~\eqref{eq:soft-scale} for ungroomed jets, and the corresponding soft scale for the groomed case $\mu_S^{\rm gr}\sim p_T(z_{\rm cut} \tau^{1+\beta}/R^\beta)^{\frac{1}{2+\beta}}$ in Eq.~\eqref{eq:soft-scale-groom} can run into the non-perturbative regime. We use profile functions~\cite{Ligeti:2008ac} in order to freeze $\mu_S$ and $\mu_S^{\rm gr}$ at 0.25~GeV in order to avoid the Landau pole. See~\cite{Kang:2018qra,Hornig:2016ahz} for more details. In order to capture non-perturbative effects we then introduce a shape function $F_i(k)$. We adopt a simple functional form for the non-perturbative shape function which only depends on a single parameter $\Omega$~\cite{Stewart:2014nna}
\bea
\label{eq:F(k)}
F_i(k) = \f{4k}{\Omega^2} \exp (-2k/\Omega)\;.
\eea
The shape function $F_i(k)$ is normalized to unity and its first moment is equal to the parameter $\Omega$:
\bea
\int_0^\infty dk\, F_i(k) = 1\;,
\qquad
\int_0^\infty dk\, k \, F_i(k) = \Omega \,. 
\eea
The subscript $i=q,g$ indicates that in principle we could have different values of $\Omega$ for quark and gluon jets. However, for our numerical calculations below, we find that a single value for $\Omega$ is sufficient. Therefore, we drop the subscript and we simply write the shape function as $F(k)$ below. For both the groomed and ungroomed jet mass distribution, we then convolve the purely perturbative result with the non-perturbative shape function. For the groomed case we have
\bea
\label{eq:shift-groom}
\f{d\sigma}{d\eta dp_T d\tau_{\rm gr}} = \int dk \,F(k) \;
\f{d\sigma^{\rm pert}}{d\eta dp_T d\tau_{\rm gr}}\left(\tau_{\rm gr} - \left(\f{k R^\beta}{p_T z_{\text{cut}}}\right)^{\f{1}{1+\beta}} \frac{k}{p_T}\right)\,.
\eea
Here $\tau_{\rm gr}$ as it is obtained from the purely perturbative result is shifted by the virtuality of the soft radiation that passes the soft drop criterion, as this mode has the smallest virtuality~\cite{Frye:2016aiz}. From Eq.~\eqref{eq:soft-scale-groom}, we identify $\mu_S^{\rm gr}\sim k$, which introduces the shift in the above formula. Analogously, for the ungroomed jet mass distribution we find~\footnote{Note that our convention for $\Omega$ here differs from~\cite{Stewart:2014nna, Liu:2014oog} by a factor of 2.}
\bea
\label{eq:shift-ungroom}
\f{d\sigma}{d\eta dp_T d\tau} = \int dk \,F(k) \;\f{d\sigma^{\rm pert}}{d\eta dp_T d\tau}\left(\tau - R \frac{k}{p_T}\right) \,.
\eea
Here the shift can be derived from Eq.~\eqref{eq:soft-scale} by identifying $\mu_{S}\sim k$. Note that also after taking into account the non-perturbative shape function, the ungroomed jet mass distribution is obtained from the groomed case by taking the limit $\beta\to\infty$. This can be seen directly from Eq.~\eqref{eq:shift-groom} which reduces to Eq.~\eqref{eq:shift-ungroom} for $\beta\to\infty$ which removes the groomer. We note that $\Omega$ characterizes the mean shift of the jet mass spectrum due to the non-perturbative effects such as hadronization and the underlying event.

Next we discuss how we estimate theoretical uncertainties. In order to estimate QCD scale uncertainties, we vary our choices of scales for each function or mode in the factorization theorem by factors of 2 around their canonical values. For the ungroomed jet mass, we have $\mu_H,~\mu_J,~\mu_C,~\mu_S$ with the canonical choices given in Eqs.~\eqref{eq:muH}, \eqref{eq:muJ}, \eqref{eq:muC} and \eqref{eq:soft-scale}, respectively. On the other hand, for groomed jets, besides $\mu_H,~\mu_J,~\mu_C$, we have two separate soft scales $\mu_{S}^{\notin{\rm gr}}$ and $\mu_{S}^{\rm gr}$, with the corresponding canonical choices given in Eqs.~\eqref{eq:notingroom} and \eqref{eq:soft-scale-groom}, respectively. We vary these scales while maintaining the relation
\bea
\f{1}{2} \leq \left.\f{\mu_i}{\mu_i^{\text{can}}}\right/\f{\mu_j}{\mu_j^{\text{can}}} \leq 2\,,
\eea
where the superscript indicates the canonical scale. Note that we choose to fix the collinear scale $\mu_C$ in terms of the soft scale $\mu_S$ for the ungroomed case and only vary them together. Thus, we have
\bea
\mu_C = \sqrt{\mu_{S} p_T R}\,.
\eea
Similarly, for the groomed case, we relate the collinear scale $\mu_C$ to the soft scale $\mu_{S}^{\rm gr}$. In addition, we fix the soft scale $\mu_{S}^{\notin{\rm gr}}$ relative to the jet scale $\mu_J$ and, thus, we only vary the two sets of scales together
\bea
\mu_C =& \left(\f{\mu_{S}^{\rm gr}}{p_T}\right)^{\f{2+\beta}{2(1+\beta)}} \left(\f{R^\beta}{z_{\rm cut}}\right)^{\f{1}{2(1+\beta)}} p_T\,,
\\
\mu_{S}^{\notin{\rm gr}} =& z_{\rm cut} \mu_J \,.
\eea

\subsection{Numerical results: the ungroomed jet mass}
For all the numerical results presented in this work we consider jets that are reclustered through the anti-$k_T$ algorithm~\cite{Cacciari:2008gp} and we use the CT14NLO PDF set~\cite{Dulat:2015mca}. We start with ungroomed jet mass distribution for the single inclusive jet production $pp\to\text{jet}+X$. In Fig.~\ref{fig:standard-jet}, we show the comparison of our theoretical calculations and the experimental data from the ATLAS collaboration which was taken at $\sqrt{s} = 7$ TeV at the LHC~\cite{ATLAS:2012am}. The shown ungroomed jet mass distributions are plotted as a function of $m_J$ and they are normalized to the inclusive jet cross section, see Eq.~(\ref{obs:ungroom}). For the experimental analysis a jet radius parameter of $R=1$ was chosen and the jets are taken into account in the rapidity range of $|\eta| < 2$. In addition, the observed jets are required to have a transverse momentum in the range of $200< p_T< 600$~GeV. The allowed jet transverse momentum range is separated into four intervals $200 < p_T < 300$ GeV, $300 < p_T < 400$ GeV, $400 < p_T < 500$ GeV, $500 < p_T < 600$ GeV which corresponds to the four panels shown in Fig.~\ref{fig:standard-jet}. The plotted experimental errors include systematic and statistical uncertainties added in quadrature. For each jet transverse momentum interval we show two theory curves, along with the results from Pythia8 simulations~\cite{Sjostrand:2014zea}. First, the dashed black lines with the yellow uncertainty bands show our purely perturbative predictions at NLL accuracy, i.e. without the non-perturbative shape function. Second, the red lines and the corresponding hatched red error bands show the theory predictions including the non-perturbative shape function as discussed in section~\ref{sec:shapefct} above. For both cases the theoretical error bands are obtained by varying the scales as discussed in section~\ref{sec:shapefct} and by taking the envelope. 
\bef
\includegraphics[width=5in]{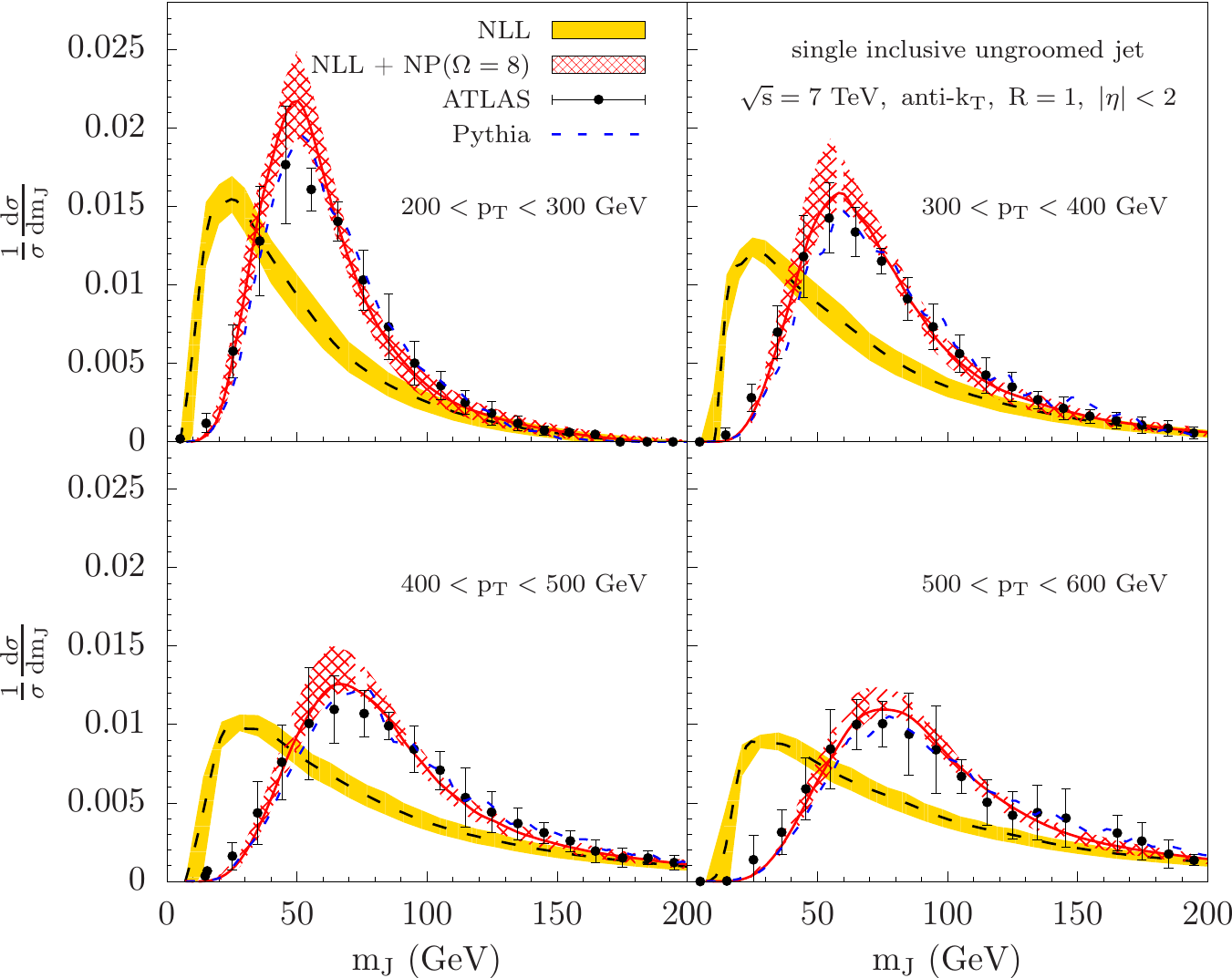} 
\caption{Comparison of our theoretical calculations and the experimental data from the ATLAS collaboration~\cite{ATLAS:2012am} taken at $\sqrt{s} = 7$ TeV. The dashed black lines and the yellow error bands show the purely perturbative results at NLL accuracy. The red lines and the red hatched error bands show the theoretical results when the non-perturbative shape function is included. The parameter of the non-perturbative shape function is chosen as $\Omega = 8$~GeV, see Eq.~\eqref{eq:F(k)}, and the distribution shown in red is obtained following Eq.~\eqref{eq:shift-ungroom}. The dashed blue lines are the results from Pythia8 simulations~\cite{Sjostrand:2014zea}. The jet rapidity is integrated over $|\eta| < 2$, and the observed jet transverse momentum is separated into four different intervals $200 < p_T < 300$ GeV, $300 < p_T < 400$ GeV, $400 < p_T < 500$ GeV, $500 < p_T < 600$ GeV which correspond to the four different panels. \label{fig:standard-jet}}
\eef

For the parameter $\Omega$ in the non-perturbative shape function we choose $\Omega = 8$~GeV which gives a very good description of the experimental data. The fact that we need such a large value for $\Omega$ reflects the fact that, as expected, the ungroomed jet mass distribution is very sensitive to non-perturbative physics such as hadronization and the underlying event etc.~\cite{Dasgupta:2012hg,Stewart:2014nna}. In fact, the position of the peak is shifted by a factor of 3 depending on the $p_T$ of the identified jets. On the other hand, the Pythia simulations that include both hadronization and underlying event contributions describe the data well, as indicated by the blue dashed curves. When grooming is taken into account the sensitivity to non-perturbative physics is expected to be significantly reduced which we confirm in the section below. Note that we did not take into account NGLs which, however, are expected to give a relatively small contribution. Nevertheless, it is remarkable that by tuning a single parameter $\Omega$ in the rather simple non-perturbative model for the shape function, the developed factorization formalism can give a very good description of the ungroomed jet mass distribution in $pp$ collisions at the LHC. One generally observes that the ungroomed jet mass distribution peaks at larger values as the $p_T$ of the identified jets is increased. This is consistent with the usual evolution picture~\cite{Konychev:2005iy}, where the larger the $p_T$ is, the longer the evolution develops. 
\bef
\includegraphics[width=5in]{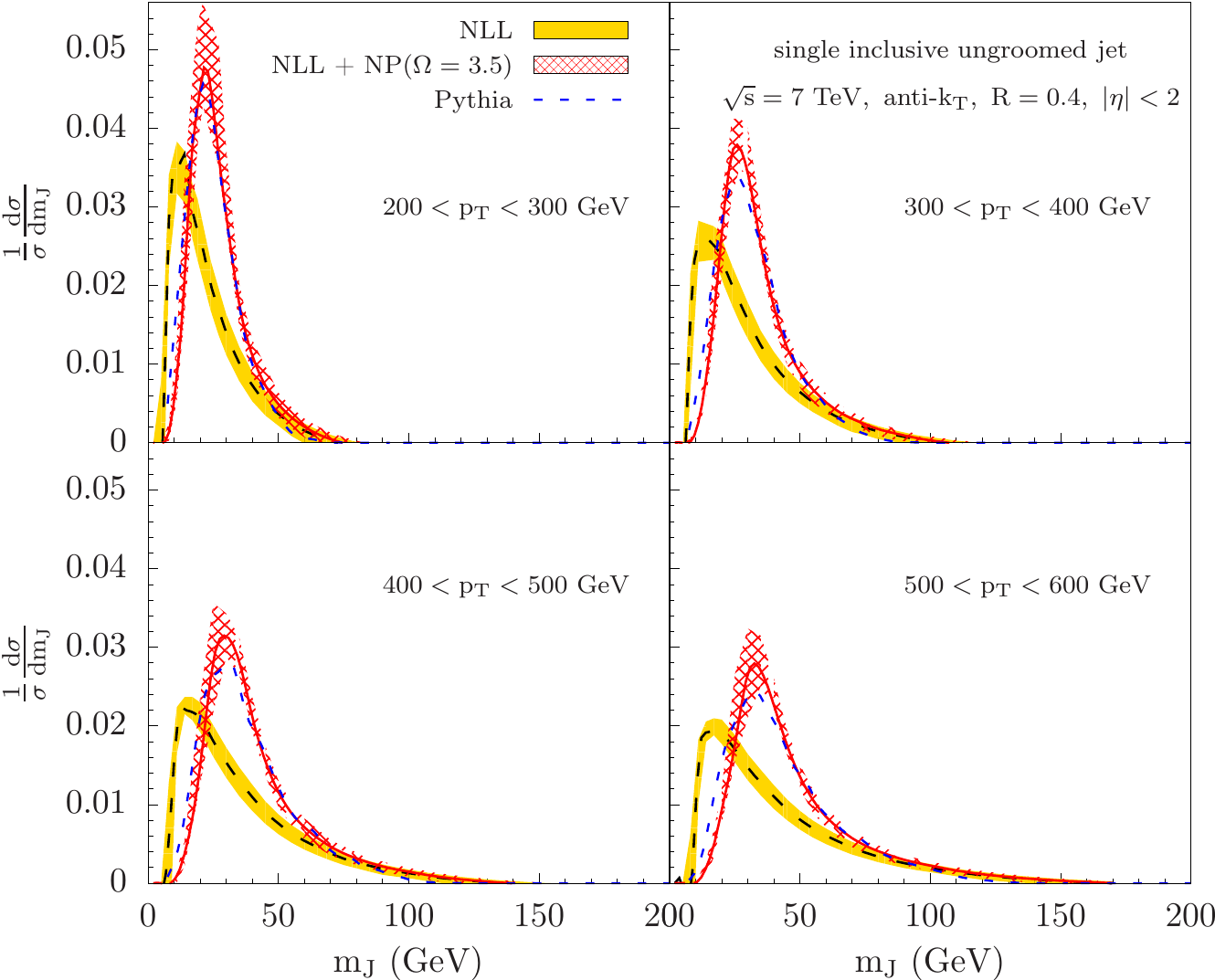}
\caption{Same as for Fig.~\ref{fig:standard-jet}, but for jets with $R=0.4$. The parameter of the non-perturbative shape function is chosen as $\Omega = 3.5$~GeV, to agree better with the Pythia8 results.}
\label{fig:standard-jet-R04} 
\eef

The fact that our factorization formalism originally derived for $R\ll 1$ works this well for such a large radius jet $R=1$ confirms earlier observation: as emphasized in Sec.~\ref{sec:ungroom-fac}, the power corrections of the form ${\mathcal O}(R^2)$ to our factorization formalism are quite small. To further test our factorization formalism and understand the non-perturbative physics, in Fig.~\ref{fig:standard-jet-R04}, we plot the jet mass distributions for jets with a smaller radius $R=0.4$ at the same kinematic regions as above. We find that the distributions are concentrated more in the small $m_J$ region compared with the larger $R$ counterparts. This is as expected, smaller $R$ leads to more collimated jets and thus smaller jet invariant mass. At the same time, we find that our perturbative results convolved with the non-perturbative shape function with a much smaller $\Omega = 3.5$~GeV than the larger $R$ case, agree very well with the Pythia simulations. This suggests that while the hadronization effect always exists, the underlying event contributions seem to be smaller for jets with smaller $R$. This is consistent with the earlier analysis~\cite{Stewart:2014nna}.

\subsection{Numerical results: the groomed jet mass \label{sec:groomed_jet}}

\bef
\includegraphics[width=\textwidth]{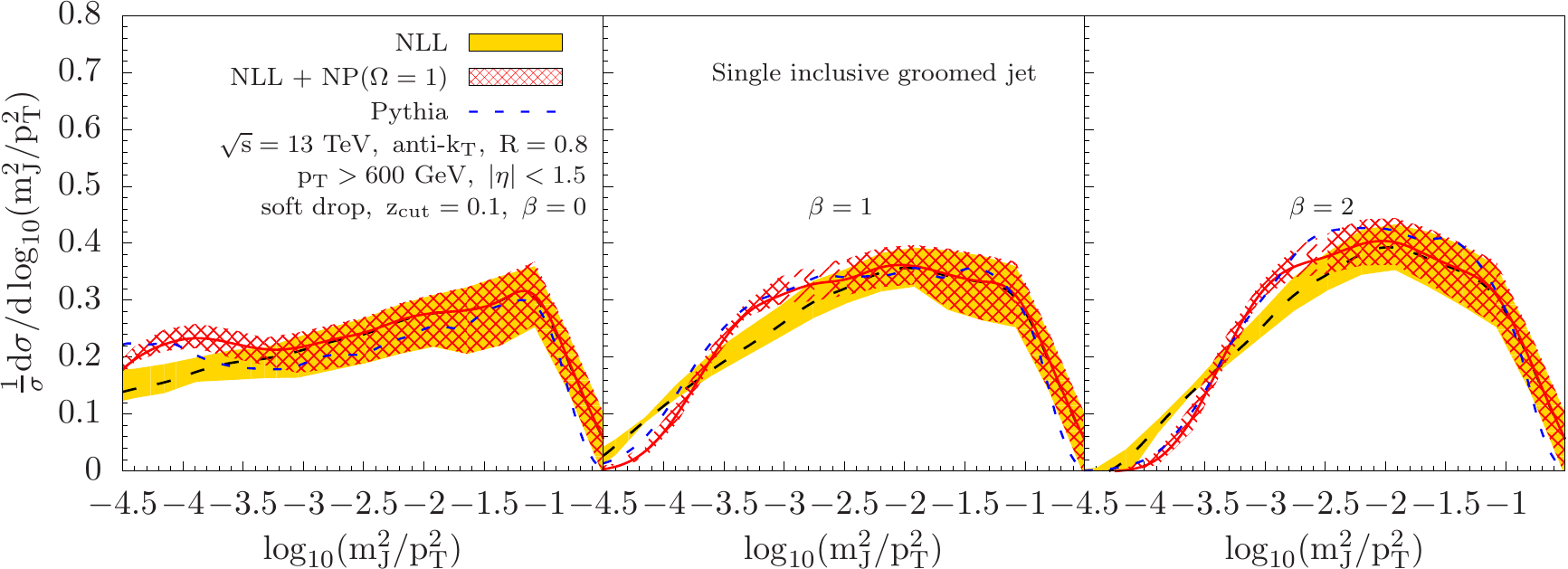} 
\caption{The theoretical predictions for the soft drop groomed jet mass distribution for single inclusive jet production $pp\to\text{jet}+X$ at $\sqrt{s} = 13$ TeV. The observed jets are reconstructed through the anti-$k_T$ algorithm with a jet radius parameter of $R=0.8$. The rapidity and transverse momentum intervals for the inclusive jet samples are chosen as $|\eta| < 1.5$ and $p_T > 600$ GeV and the soft threshold parameter is $z_{\rm cut} = 0.1$. The soft drop groomed jet mass distribution is normalized to the corresponding inclusive jet cross section and plotted as a function of $\log_{10}(m_{J,{\rm gr}}^2/p_T^2)$ for $\beta = 0$ (left), $\beta=1$ (middle), and $\beta = 2$ (right). The dashed black lines and yellow error bands show the purely perturbative NLL results, while the red lines and the red hatched bands are the NLL results but including the non-perturbative shape function according to Eq.~\eqref{eq:shift-groom}. We choose $\Omega = 1$~GeV for the parameter of non-perturbative shape function. The dashed blue lines are from Pythia simulations.
\label{fig:inclusive-groom}}
\eef

In this section we are now going to present numerical results for the soft drop groomed jet mass distribution for single inclusive jet production $pp\to\text{jet}+X$ at the LHC. Unfortunately, there is currently no data available for inclusive jet production that would allow for a direct one-to-one comparison to our theoretical results. In~\cite{CMS:2017xdn}, the CMS collaboration presented preliminary results for the groomed jet mass distribution for inclusive jet production for both Pb-Pb and $pp$ collisions at $\sqrt{s}=5.02$~TeV but the $pp$ baseline is smeared to allow for a better comparison to the heavy-ion data. Nevertheless, we expect that such an analysis of LHC data is feasible and will become available in the near future. With this in mind, we present our predictions for $\sqrt{s}=13$~TeV at the LHC. As an example, we assume that jets are reconstructed using the anti-$k_T$ algorithm with a jet radius parameter of $R=0.8$. We choose the following jet transverse momentum and rapidity intervals for the inclusive jet sample: $|\eta| < 1.5$ and $p_T > 600$ GeV. For the soft threshold parameter of the soft drop grooming algorithm, we choose $z_{\rm cut} = 0.1$. 

In Fig.~\ref{fig:inclusive-groom}, we show the soft drop groomed jet mass distributions normalized by the corresponding inclusive jet cross sections as a function of $\log_{10}(m_{J,{\rm gr}}^2/p_T^2)$ for three different values of the angular exponent: $\beta = 0$ (left), $\beta=1$ (middle), and $\beta = 2$ (right). Similar to Fig.~\ref{fig:standard-jet}, the dashed black lines and the corresponding yellow error bands show our purely perturbative results at NLL accuracy. The red lines and the red hatched bands show the result when the non-perturbative shape function is included where we follow the prescription in Eq.~\eqref{eq:shift-groom}. We choose the parameter of the non-perturbative shape function as $\Omega = 1$ GeV to illustrate the impact of non-perturbative physics effects. Finally, the dashed blue lines are from Pythia simulations. We find that the numerical results from our factorization formalism with $\Omega = 1$ GeV agree well with the Pythia results, for a relatively large jet radius $R=0.8$. This much reduced parameter $\Omega$ compared to the ungroomed cases indicates that the groomed jet mass distributions have a much smaller sensitivity to the nonperturbative physics. The fact that $\Omega = 1$ GeV is around the size of a typical hadron mass implies that the nonperturbative contributions come mainly from hadronization. 
\bef
\includegraphics[width=\textwidth]{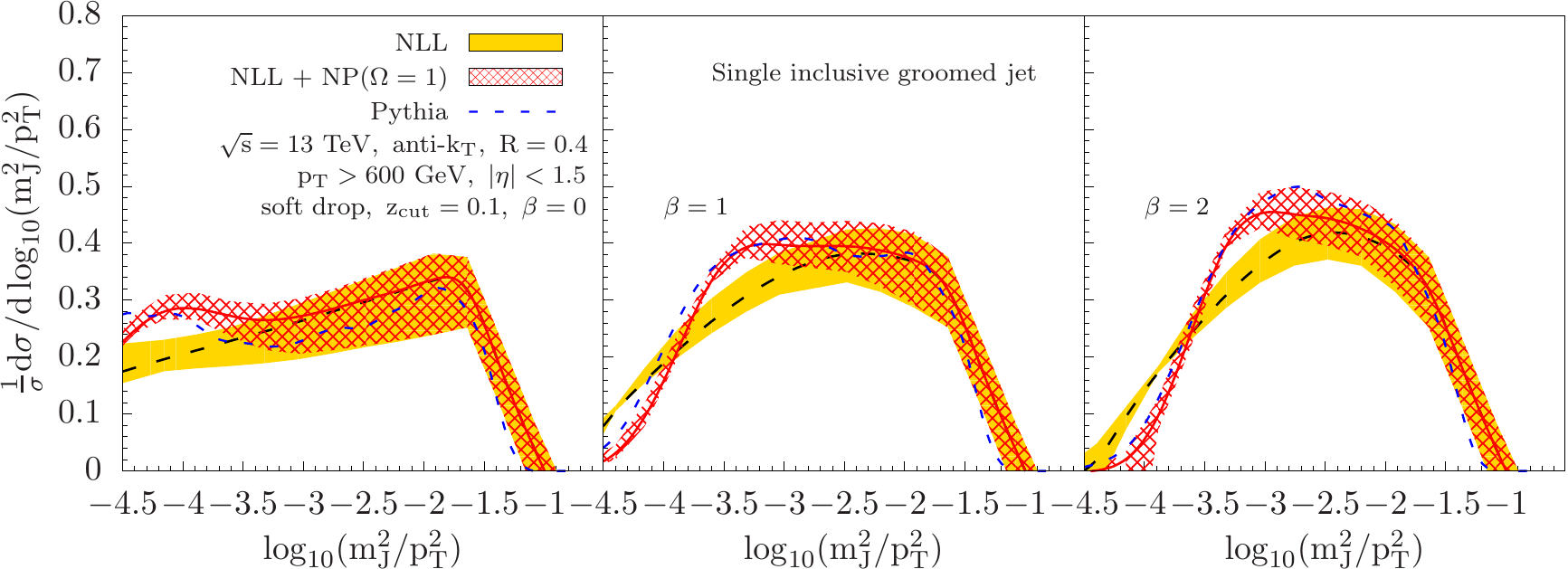} 
\caption{Same as for Fig.~\ref{fig:inclusive-groom}, but for jets with $R=0.4$. The parameter of the non-perturbative shape function is chosen as $\Omega = 1$~GeV, to agree better with the Pythia8 results.}
\label{fig:inclusive-groom-R04}
\eef

To further test our factorization formalism for groomed jet substructure and to understand the nonperturbative physics, in Fig.~\ref{fig:inclusive-groom-R04}, we plot the groomed jet mass distributions for jets with a smaller radius $R=0.4$ at the same kinematic regions as in Fig.~\ref{fig:inclusive-groom}. We find that the same parameter $\Omega=1$~GeV leads to a good agreement between our numerical results and the Pythia simulations. This strongly suggests that the underlying event contributions are much reduced for the groomed jet mass distribution, and the main nonperturbative physics comes from hadronization. 

As discussed above, the ungroomed jet mass distribution should be recovered from the groomed case by taking the limit $\beta\to\infty$. In section~\ref{sec:consistency} we discussed this transition at the level of the analytical perturbative results. Here, we study the $\beta\to\infty$ limit numerically. In Fig.~\ref{fig:betalimit}, we plot the groomed jet mass distribution for different values of the angular exponent in the range of $\beta=0$ to 4 (dashed lines) as well as the ungroomed result (solid blue). Note that we only show the purely perturbative results here in order to better illustrate how the groomed results converge to the ungroomed jet mass distribution when $\beta$ is increased. If we instead have included non-perturbative shape function, then $\Omega$ would have to be adjusted when taking the limit $\beta\to\infty$. Note that here we plot both the groomed and ungroomed results as a function of $\log_{10}(m_{J,{\rm gr}}^2/p_T^2)$ as in Fig.~\ref{fig:inclusive-groom} instead of $m_J$ used in Fig.~\ref{fig:standard-jet}. For a stronger grooming procedure (smaller values of $\beta$), the jet mass distribution gets flatter and shifted toward smaller values. This is expected intuitively as it becomes more likely to observe smaller values of the jet mass after the grooming procedure which removes soft wide-angle radiation from the jet. As expected a smooth transition between the groomed and ungroomed case can be observed for $\beta\to\infty$. This feature of the jet mass distributions can be particularly useful in order to understand the impact of grooming in heavy-ion collisions, see for example~\cite{Acharya:2017goa,CMS:2017xdn}. 

The groomed jet mass distributions for different values of $\beta$ all become very similar at the transition point $\tau_{\rm gr} = m_{J,{\rm gr}}^2/p_T^2 = z_{{\rm cut}} R^2$. This can also be seen from the values of the soft scales. By identifying $ \tau_{\rm gr} = z_{\rm cut} R^2$, we find from Eqs.~\eqref{eq:notingroom} and~\eqref{eq:soft-scale-groom}, 
\bea
\mu_S^{\rm gr}|_{{\tau_{\rm gr}} =  z_{\rm cut} R^2}=\mu_{S}^{\notin{\rm gr}}|_{{\tau_{\rm gr}}=  z_{{\rm cut}} R^2} =  \frac{p_T \tau_{\rm gr} }{R}= \mu_S \,,
\eea
which makes the scales of the soft functions in the groomed case to be identical to the scale of the soft function for the ungroomed case, see Eq.~\eqref{eq:soft-scale}. This makes the evolution factors identical independent of $\beta$ values and whether there is a grooming or not, and $\beta$ dependence only enters from the renormalized expressions of the soft functions at the fixed-order. Therefore, although in reality the perturbative results do not all intersect exactly at $\tau = z_{{\rm cut}} R^2$, they become very similar at $\tau = z_{{\rm cut}} R^2$ as can be seen from Fig.~\ref{fig:betalimit}. At larger values, the grooming does not play a role and the ungroomed jet mass distribution is recovered. See the discussion in section~\ref{sec:scfacgr} and the Appendix~\ref{app:softpass}. 

\bef
\includegraphics[width=0.5\textwidth]{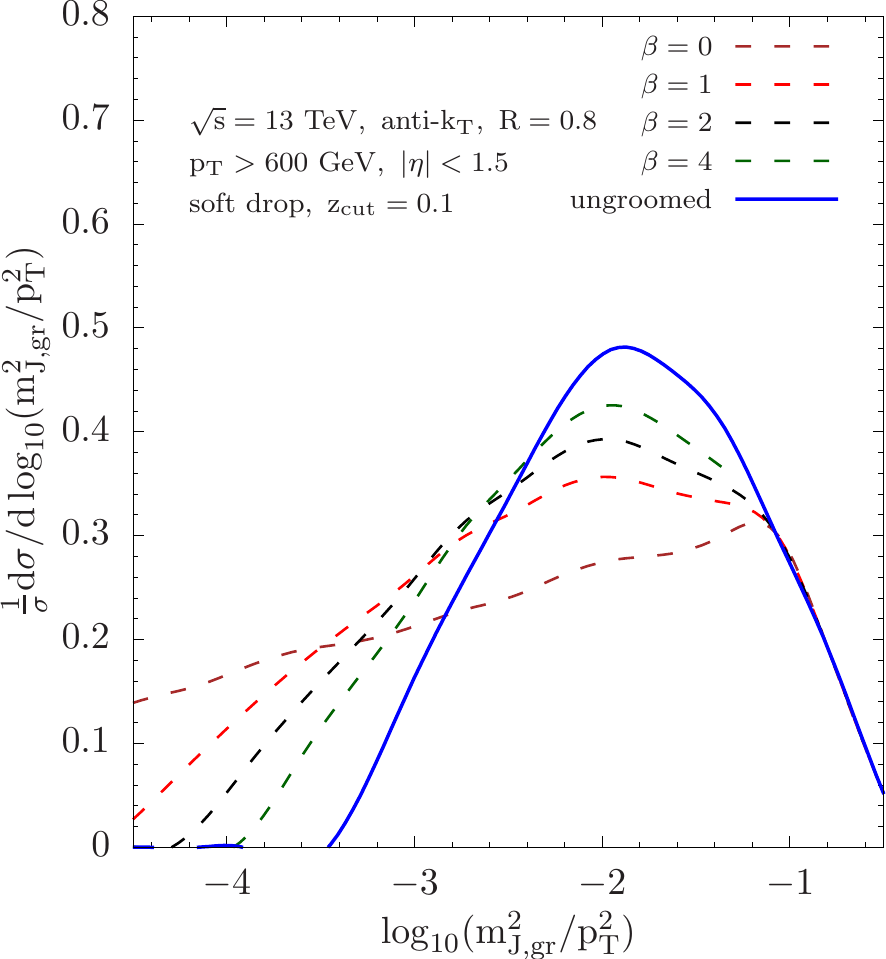} 
\caption{The ungroomed (solid blue) and groomed (dashed lines) jet mass distributions for different values of the angular exponent $\beta$. The same kinematical setup is used as in Fig.~\ref{fig:inclusive-groom}. We only show the purely perturbative results plotted as a function of $\log_{10}(m_{J,{\rm gr}}^2/p_T^2)$. In the limit $\beta\to\infty$, the ungroomed distribution is recovered from the groomed case. \label{fig:betalimit}}
\eef

Recently the ATLAS collaboration reported on a measurement of the soft drop groomed jet mass distribution in~\cite{Aaboud:2017qwh}. A similar analysis was performed by CMS in~\cite{CMS:2017tdn}. The measurement is performed in an inclusive way in the sense that no additional cuts are imposed on the hadronic activity outside the signal jets. However, additional cuts are imposed on the observed jet transverse momenta which unfortunately hinders a direct one-to-one comparison with the inclusive jet production framework developed in this work. The details of the analysis are as follows. Events are taken into account that have at least two jets and the leading jet is required to have a transverse momentum of $p_{T1}>600$~GeV. In addition, the two leading $p_T$-ordered jets are required to satisfy $p_{T,1} / p_{T,2} <  1.5$. Since the two leading jets are required to have a similar transverse momentum, this additional requirement effectively enforces a di-jet configuration. Events with additional energetic jets are thus removed. The two leading jets are then included in the soft drop jet mass measurement. Furthermore, the $\eta$ of the thus obtained jet samples is restricted to $|\eta| < 1.5$.

The ATLAS results for the groomed jet mass distribution are then plotted as
\bea 
\label{logdist}
\f{1}{\sigma_{\text{resum}}} \f{d\sigma}{d\log_{10} \tau_{\rm gr}}\,.
\eea
Here $\sigma_{\rm resum}$ in the denominator is the integrated cross section measured in the so-called a ``resummation'' region~\cite{Aaboud:2017qwh} and it is defined as
\bea
\label{eq:resum-exp}
\sigma_{\rm resum} \equiv  \int_{-3.7}^{-1.7}\f{d\sigma}{d\log_{10} \tau_{\rm gr}} d\log_{10}\tau_{\rm gr}\,.
\eea
The factorization formalism developed in this work is for single inclusive jet production, which is strictly speaking not compatible with the ATLAS measurement. However, most of the events are indeed di-jet configurations when the jet $p_T$ is very large~\cite{Liu:2017pbb}. The additional production of a third jet with a very large transverse momentum is suppressed by an additional power of $\alpha_s(p_T)\ll 1$. One can thus expect that the qualitative features of the soft drop jet mass distribution as measured by ATLAS are nevertheless correctly described by the factorization formalism presented in this work for $pp\to\text{jet}+X$. However, we note that high precision jet substructure studies require a direct one-to-one correspondence between the experimental measurement and the theoretical calculations.

\bef
\includegraphics[width=\textwidth]{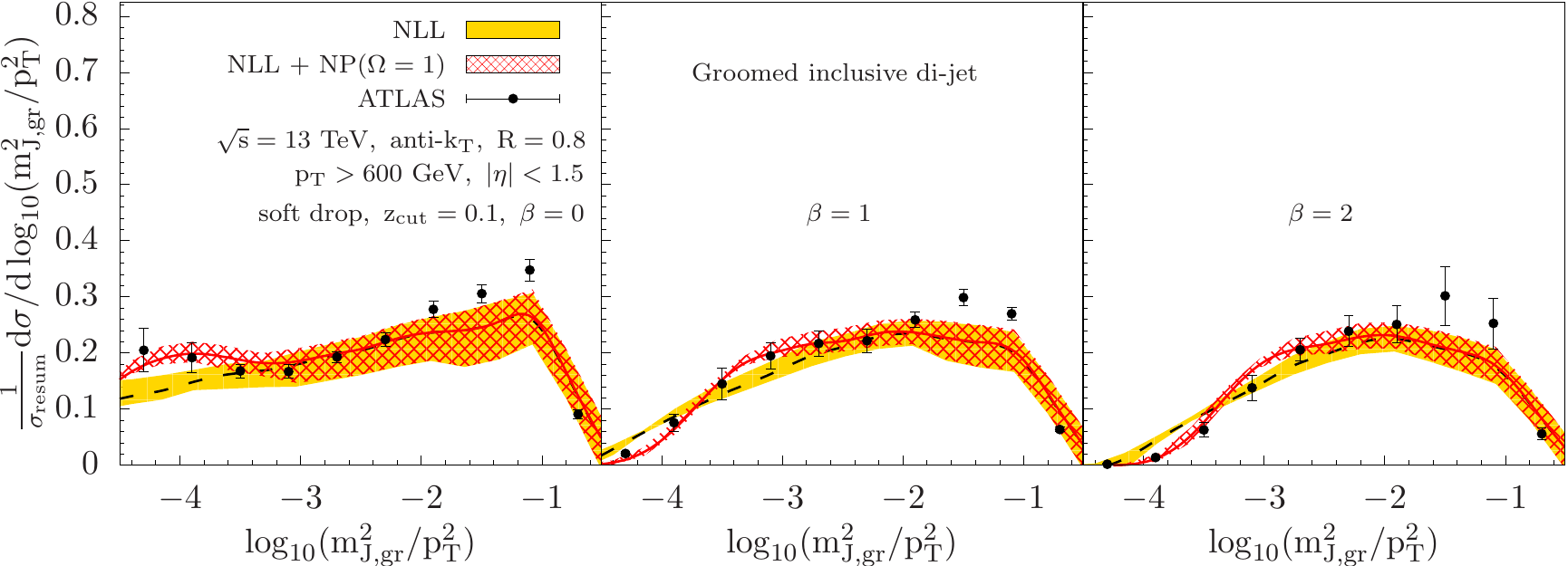} 
\caption{Comparison of our theoretical calculations and the experimental data from the ATLAS collaboration~\cite{Aaboud:2017qwh} for the soft drop groomed jet mass distribution in $pp$ collisions at $\sqrt{s} = 13$ TeV at the LHC. See text for a more detailed discussion. \label{fig:compare-groom}}
\eef

Instead of normalizing the soft drop jet mass distribution by the inclusive cross section as shown in Fig.~\ref{fig:inclusive-groom}, we now adopt the normalization used by ATLAS and divide by $\sigma_{\rm resum}$.  We thus follow Eq.~\eqref{eq:resum-exp} and integrate our results over the range of $-3.7<\log_{10}\tau_{\rm gr}<-1.7$. In Fig.~\ref{fig:compare-groom}, the ATLAS data for the soft drop groomed jet mass distribution is shown where both systematic and statistical errors are included. The data is plotted as a function of $\log_{10}(m_{J,{\rm gr}}^2/p_T^2)$ for three different values of the angular exponent used in the analysis: $\beta = 0$ (left), $\beta = 1$ (middle), and $\beta = 2$ (right). In addition, we show the theoretical results using the factorization formalism developed in this work for the groomed jet mass distribution. As in Fig.~\ref{fig:inclusive-groom}, the dashed line and the corresponding yellow error band are the purely perturbative results at NLL accuracy, while the red line and the red hatched band are NLL results convolved with the non-perturbative shape function. Again we choose $\Omega = 1$~GeV which gives a very good description of the experimental data in the resummation region. In this region, the factorization formalism developed here is expected to work very well. By including the non-perturbative shape function, the agreement with the data can also be achieved in the very small jet mass region. In the very large jet mass region, we would have to include a matching to fixed order calculations. In general, it is possible to include such a matching in our formalism which, however, is beyond the scope of this work and will be addressed in the future.

 The soft drop grooming procedure is designed to eliminate the sensitivity to the underlying event contribution. This is confirmed by our numerical results for the different jet mass distributions. Note that we consistently treat non-perturbative effects for both the groomed and the ungroomed case by using the same shape functions. We find that the non-perturbative parameter $\Omega = 1$~GeV is much smaller for the groomed case than for the ungroomed jet mass distribution where we had to use $\Omega=8$~GeV in order to find a good agreement with the data. Note that our calculations are performed at the parton level whereas the experimental results are unfolded at the hadron level. Therefore, a remaining non-perturbative correction needs to be taken into account also in the groomed case. However, this remaining hadronization correction is expected to be small since it should be at the order of $\Lambda_{\rm QCD}$. This expectation agrees with our observation that $\Omega=1$~GeV is sufficient in order to obtain a good agreement with the experimental data.

\section{Summary and outlook \label{sec:summary}}
In this work, we studied the jet mass distribution for the single inclusive jet production at the LHC, fully differential in the kinematics of the signal jet. We considered both the ungroomed and soft drop groomed mass distributions.

We derived the corresponding factorization formalisms from first principles in perturbative QCD. We studied the connections and differences between the factorization theorems for the groomed and ungroomed case, and we computed all the necessary components to NLO. By solving the associated renormalization group equations, we are able to perform the joint resummation at the NLL accuracy, for logarithms in both the small jet radius parameter $R$ and the small jet mass $m_J$. For the soft drop groomed jet mass distribution, an additional resummation of the logarithms in the soft threshold parameter $z_{\rm cut}$ has also been achieved. In this sense, we realized for the first time a complete description of the groomed inclusive jet mass distribution where all relevant logarithms have been resummed at NLL accuracy. The complete resummation of logarithms in $z_{\rm cut}$ allows us to reliably determine the absolute normalization of groomed jet observables. In addition, the derived factorization theorem allows for systematically including NGLs in the future. Being able to completely resum logarithms in the soft threshold parameter $z_{\rm cut}$ will enable a comparison of theory calculations and data where significantly smaller values are chosen for $z_{\rm cut}$ which can be advantageous in some situations. In addition, the resummation of single logarithms in the jet size parameter is particularly useful for jets measured in heavy-ion collisions where typically a rather small jet radius parameter is chosen.

It is important to realize that the developed hard collinear factorization formalism established in this work enables us to compute the relative contribution of jets that are initiated by either quarks or gluons. Such a relative fraction of quark and gluon jets in the sample can be determined order by order in the perturbation theory through the computation of the hard functions $H_{ab}^c$. For the current study, we have used the NLO hard functions $H_{ab}^c$ and thus the relative ratios are computed to NLO accuracy. This is apparently an advantage of our factorization formalism for single inclusive jet production. Our formalism allows for the extension to tagged jets observed in inclusive processes like $pp\to Z+\text{jet}+X$ which we are planning to address in forthcoming work.

We further presented numerical results for the ungroomed inclusive jet mass distributions at the LHC,  with the experimental kinematic cuts fully taken into account. For the groomed jet mass spectrum, a direct one-to-one comparison with LHC data is currently not feasible as there are no soft drop groomed jet mass measurements available for inclusive jet production. Instead, we compared our predictions with the groomed jet mass distribution measured in high $p_T$ di-jet events, based on the observation that the inclusive cross section is dominated by di-jet configurations at large jet transverse momenta.  In general, we found that our theoretical calculations lead to a very good description of the experimental data in the regions where the factorization theorems hold. Given the success of our formalism for inclusive jet production and the advantages in statistics, we suggest that the soft drop groomed jet mass measurement should also be performed using inclusive jet samples in the future. 

To further extend the region of validity of our formalism, a further matching of the NLL results to the full NLO calculation is required which will be left for future work. The full NLO calculations can be achieved using nlojet++~\cite{Nagy:2003tz}. Computations beyond NLL accuracy for single inclusive jet samples are also possible but would require the calculation of the hard functions $H_{ab}^c$ for producing a single inclusive parton to NNLO which is the un-renormalized partonic cross section for producing a parton (not a jet) in the final state. This task is challenging, but recent studies for the single inclusive jet cross section at NNLO~\cite{Currie:2016bfm} make it very promising in the near future. We expect that the framework developed in this work can be directly generalized to study other groomed jet substructure observables.

\acknowledgments
We thank P.~Jacobs, I.~Moult, B.~Nachman, D.~Neill, M.~Ploskon, A.~Sickles, G.~Sterman, I.~Stewart, V.~Theeuwes and F.~Yuan for helpful discussions. Z.K. is supported by the National Science Foundation under Grant No.~PHY-1720486. K.L. is supported by the National Science Foundation under Grants No.~PHY-1316617 and No.~PHY-1620628. X.L. is supported by the National Natural Science Foundation of China under Grant No.~11775023 and the Fundamental Research Funds for the Central Universities. F.R. is supported by the Department of Energy under Contract No.~DE-AC0205CH11231, and the LDRD Program of Lawrence Berkeley National Laboratory.

\appendix
\section{Soft functions for the groomed jet mass \label{sec:appendixA}}
In this Appendix we provide some details of the calculation of the soft functions in the presence of soft drop grooming. We start with the soft drop criterion in Eq.~\eqref{eq:SD} which is boost-invariant. Therefore we can choose to perform our calculations in a reference frame in which the jet has no transverse momentum component relative to the jet direction. In this frame, the four-momentum of the jet can be written as $\ell^\mu = (\ell^-=\omega_J, \ell^+, 0_\perp)$ with jet energy $E_J \approx \omega_J/2$. In such a reference frame the jet energy is given by the observed jet transverse momentum in the center-of-mass frame, i.e. we have $E_J = p_T$. Thus, the soft drop criterion in Eq.~\eqref{eq:SD} can be written for soft radiation as
\bea
\frac{k^0}{E_J + k^0} \approx \frac{k^0}{\omega_J/2} > z_{\rm cut} \left(\frac{\theta_{ij}}{R}\right)^\beta\,,
\label{eq:SD0}
\eea
where $k$ denotes the soft momentum. Here $\theta_{ij}$ is the angle between the soft particle and the jet axis, which can be determined from~\cite{Ellis:2010rwa} to be
\bea
\tan^2\left(\frac{\theta_{ij}}{2}\right) = \frac{k^+}{k^-}\,. 
\eea
We may thus rewrite Eq.~\eqref{eq:SD0} as
\bea
k^- + k^+ > z_{\rm cut} \,\omega_J \left(\frac{4}{R^2}\frac{k^+}{k^-}\right)^{\beta/2},
\eea
where we have used $\tan(\theta/2)\approx \theta/2$ for collimated jets ($R\ll 1$). 

\subsection{Soft radiation that fails the soft drop criterion}
First we provide the details of the calculation for the soft functions $S_i^{\notin{\rm gr}} (p_T, R, \mu; z_{\rm cut}, \beta)$, which describe soft radiation that fails the soft drop criterion. Since this radiation fails the soft drop criterion, it is removed from the jet and, thus, does not contribute to the observed groomed jet mass. In this case, the soft momentum $k$ within the jet will satisfy the following constraints
\bea\label{eq:constraint1}
k^-  + k^+ &< z_{\rm cut} \,\omega_J \left(\frac{4}{R^2}\frac{k^+}{k^-}\right)^{\beta/2}\,,
\\
& \frac{k^+}{k^-} < \left(\frac{R}{2}\right)^2\,.\label{eq:constraint2}
\eea
Here the first inequality states that the soft radiation fails the soft drop criterion, while the second one is the constraint on the soft momentum $k$ within the jet for the anti-$k_T$ algorithm. With that, the non-vanishing contribution to the NLO correction for the soft functions $S_i^{\notin{\rm gr}}$ within $\overline{\rm MS}$ scheme are given by
\bea
S_i^{\notin{\rm gr}} (p_T, R, \mu; z_{\rm cut}, \beta) =& 32 \pi^2 \alpha_s C_i \left(\frac{\mu^2 e^{\gamma_E}}{4\pi}\right)^\epsilon \int \frac{d^{n}k}{(2\pi)^n}\delta(k^2) \frac{1}{k^+ k^-} 
\nnu
&\times
\Theta\left(k^-  + k^+ < z_{\rm cut} \,\omega_J \left(\frac{4}{R^2}\frac{k^+}{k^-}\right)^{\beta/2}\right) \Theta\left(\frac{k^+}{k^-} < \left(\frac{R}{2}\right)^2\right)\,,
\eea
where the space-time dimensions are given by $n=4-2\epsilon$. We use the notation $C_i = C_{F, A}$ for the quark and gluon soft functions, respectively. Changing integration variables from $k^+, ~k^-$ to $x,~y$ where 
\bea
x = k^+ + k^-\,,
\qquad
y = \frac{k^+}{k^-}\,, 
\eea
we find
\bea
S_i^{\notin{\rm gr}} (p_T, R, \mu; z_{\rm cut}, \beta)  =&\, \frac{\alpha_s}{2\pi} C_i \frac{(\mu^2 e^{\gamma_E})^\epsilon}{\Gamma(1-\epsilon)} \int_0^{\infty} \frac{dx}{x^{1+2\epsilon}} \int_0^\infty dy \frac{(1+y)^{2\epsilon}}{y^{1+\epsilon}}
\nnu
&\times  \Theta\left(x < z_{\rm cut} \,\omega_J \left(\frac{4}{R^2}y\right)^{\beta/2}\right) \Theta\left(y < \left(\frac{R}{2}\right)^2\right)
\nnu
=&\,
\frac{\alpha_s}{2\pi} \frac{C_i }{1+\beta} \frac{e^{\epsilon \gamma_E}}{\Gamma(1-\epsilon)}\frac{1}{\epsilon^2} \left(  \frac{\mu^2}{z_{\rm cut}^2p_T^2R^2}\right)^\epsilon + {\cal O}(R^2)\,.
\eea
Here we neglected the power corrections of the form ${\cal O}(R^2)$. After expanding in powers of $\epsilon$, we obtain
\bea
S_i^{\notin{\rm gr}} (p_T, R, \mu; z_{\rm cut}, \beta)  = \frac{\alpha_s}{2\pi}  \frac{C_i}{1+\beta}  
\left[\frac{1}{\epsilon^2}+\frac{1}{\epsilon}\ln\left(\frac{\mu^2}{z_{\rm cut}^2p_T^2R^2}\right)
+\frac{1}{2}\ln^2\left(\frac{\mu^2}{z_{\rm cut}^2p_T^2R^2}\right) - \frac{\pi^2}{12}
\right].
\eea
Next we consider the renormalization of the above soft functions. The bare and renormalized quantities are related multiplicatively as follows
\bea
S_{i,\rm bare}^{\notin{\rm gr}} (p_T, R; z_{\rm cut}, \beta)  = Z_{S_i}^{\notin{\rm gr}}(p_T, R, \mu; z_{\rm cut}, \beta) \, S_i^{\notin{\rm gr}} (p_T, R, \mu; z_{\rm cut}, \beta)\,,
\eea
where the renormalization constants $Z_{S_i}^{\notin{\rm gr}}$ are given by
\bea
Z_{S_i}^{\notin{\rm gr}}(p_T, R, \mu; z_{\rm cut}, \beta) = 1 + \frac{\alpha_s}{2\pi}  \frac{C_i}{1+\beta}  
\left[\frac{1}{\epsilon^2}+\frac{1}{\epsilon}\ln\left(\frac{\mu^2}{z_{\rm cut}^2p_T^2R^2}\right)\right]\,.
\eea
We thus obtain the renormalized soft functions as
\bea
S_i^{\notin{\rm gr}} (p_T, R, \mu; z_{\rm cut}, \beta) = 1+ \frac{\alpha_s}{2\pi} \frac{C_i}{1+\beta} \left[\frac{1}{2}\ln^2\left(\frac{\mu^2}{z_{\rm cut}^2 p_T^2 R^2}\right) - \frac{\pi^2}{12}\right]\,.
\eea
The associated RG equations take the following form
\bea
\mu \f{d}{d\mu} S_i^{\notin{\rm gr}} (p_T, R, \mu; z_{\rm cut}, \beta) = \gamma_{S_i}^{\notin{\rm gr}}(p_T, R, \mu; z_{\rm cut}, \beta)\, S_i^{\notin{\rm gr}} (p_T, R, \mu; z_{\rm cut}, \beta)\,,
\eea
where the anomalous dimensions $\gamma_{S_i}^{\notin{\rm gr}}$ are given by
\bea
\gamma_{S_i}^{\notin{\rm gr}}(p_T, R, \mu; z_{\rm cut}, \beta)  = \frac{\alpha_s}{\pi} \frac{C_i}{1+\beta} \ln\left(\frac{\mu^2}{z_{\rm cut}^2 p_T^2 R^2}\right)\,.
\eea

\subsection{Soft radiation that passes the soft drop criterion \label{app:softpass}}
We are now going to provide the details of the calculation of the soft functions $S_i^{\rm gr}(\tau, p_T, R, \mu; z_{\rm cut}, \beta)$, which describe soft radiation that passes the soft drop criterion. Since the associated soft particles pass the soft drop criterion, they remain in the groomed jet and thus contribute to the groomed jet mass. Therefore, the soft functions here depend on $\tau$. The NLO corrections to the soft functions $S_i^{\rm gr}(\tau, p_T, R, \mu; z_{\rm cut}, \beta)$ can be written as
\bea\label{eq:softpass}
S_i^{\rm gr}(\tau, p_T, R, \mu; z_{\rm cut}, \beta) =& 32 \pi^2 \alpha_s C_i \left(\frac{\mu^2 e^{\gamma_E}}{4\pi}\right)^\epsilon \int \frac{d^{n}k}{(2\pi)^n}\delta(k^2) \frac{1}{k^+ k^-} \delta\left(\tau - \frac{4k^+}{\omega_J}\right)
\nnu
&\times
\Theta\left(k^-  + k^+ > z_{\rm cut} \,\omega_J \left(\frac{4}{R^2}\frac{k^+}{k^-}\right)^{\beta/2}\right) \Theta\left(\frac{k^+}{k^-} < \left(\frac{R}{2}\right)^2\right)\,.
\eea
Note that the delta function $\delta(\tau- 4k^+/\omega_J)$ in the first line states the fact that the soft radiation here contributes to the jet mass via $m_s^2 = \omega_J k^+$ and $\tau = 4m_s^2/\omega_J^2$. The first $\Theta$-function in the second line is the soft drop criterion, and the second $\Theta$-function is again due to the jet algorithm constraint, see Eqs.~(\ref{eq:constraint1}) and~(\ref{eq:constraint2}) above. Both theta functions give constraints on the integration variables $k^{\pm}$ and in the following we determine which one sets a more stringent constraint on the corresponding integration regions. To proceed, we first note that $k^+ \ll k^-$ for $k^+ = \omega_J \tau/4$. This holds true in the kinematic region we are interested in $\tau/R^2 \ll z_{\rm cut}\ll 1$. The soft drop criterion can thus be simplified as
\bea
k^-  > z_{\rm cut} \,\omega_J \left(\frac{4}{R^2}\frac{k^+}{k^-}\right)^{\beta/2}\,.
\label{eq:SD-cs}
\eea
For $\tau/R^2 \ll z_{\rm cut}$, one finds that soft drop criterion obtained here in Eq.~\eqref{eq:SD-cs} is always the stronger constraint on the soft radiation than the jet algorithm constraint. Therefore, as long as Eq.~\eqref{eq:SD-cs} is satisfied, we can remove the jet algorithm constraint. By making use of these considerations, we obtain the following result up to NLO
\bea
S_i^{\rm gr}(\tau, p_T, R, \mu; z_{\rm cut}, \beta) =&\,\delta(\tau) + \frac{\alpha_s}{\pi} C_i\Bigg[
-\frac{2+\beta}{2(1+\beta)} \frac{1}{\epsilon^2} \delta(\tau) + \frac{1}{\epsilon} A \left(\frac{1}{A\tau}\right)_+ 
\nnu
&
+\frac{\pi^2}{24}\frac{2+\beta}{1+\beta} \delta(\tau) - \frac{2(1+\beta)}{2+\beta} A \left(\frac{\ln\left(A\tau\right)}{A\tau} \right)_+
\Bigg]\,,
\eea
where the factor $A$ is given by
\bea
A =\left[ \left(\frac{z_{\rm cut}}{R^\beta}\right)^{\frac{1}{2+\beta}} \frac{p_T}{\mu}\right]^{\frac{2+\beta}{1+\beta}}\,.
\eea
The bare and renormalized soft functions are related through a convolution relation as
\bea
S_{i, \rm bare}^{\rm gr}(\tau, p_T, R; z_{\rm cut}, \beta)  = \int d\tau' Z_{S_i}^{\rm gr}(\tau - \tau', p_T, R, \mu; z_{\rm cut}, \beta) S_i^{\rm gr}(\tau', p_T, R, \mu; z_{\rm cut}, \beta)\,.
\eea
The renormalization constants $Z_{S_i}^{\rm gr}$ are given by
\bea
Z_{S_i}^{\rm gr}(\tau, p_T, R, \mu; z_{\rm cut}, \beta) = \delta(\tau) + \frac{\alpha_s}{\pi} C_i\Bigg[
-\frac{2+\beta}{2(1+\beta)} \frac{1}{\epsilon^2} \delta(\tau) + \frac{1}{\epsilon} A \left(\frac{1}{A\tau}\right)_+ \Bigg]\,, 
\eea
and the renormalized soft functions are thus
\bea
S_i^{\rm gr}(\tau, p_T, R, \mu; z_{\rm cut}, \beta) = \delta(\tau) + \frac{\alpha_s}{\pi} C_i \left[
\frac{\pi^2}{24}\frac{2+\beta}{1+\beta} \delta(\tau) - \frac{2(1+\beta)}{2+\beta} A \left(\frac{\ln\left(A\tau\right)}{A\tau} \right)_+ \right]\,.
\eea
The corresponding RG equations are given by
\bea
\mu \f{d}{d\mu} S_i^{\rm gr}(\tau, p_T, R, \mu; z_{\rm cut}, \beta) =& \int d\tau' \; \gamma_{S_i}^{\rm gr}(\tau - \tau',p_T,R,\mu; z_{\rm cut}, \beta) 
\nnu
&\times
S_i^{\rm gr}(\tau', p_T, R, \mu; z_{\rm cut}, \beta)\,,
\eea
with the anomalous dimensions
\bea
\gamma_{S_i}^{\rm gr}(\tau,p_T,R, \mu; z_{\rm cut}, \beta) = \f{2\alpha_s C_i}{\pi}\left[\left(\f{1}{\tau}\right)_++ \ln(A)
\delta(\tau) \right] \,.
\eea
Next we consider the kinematic region with the parametric scaling $\tau/R^2\sim z_{\rm cut}\ll 1$. We are going to find that there is a transition point at $\tau/R^2=z_{\rm cut}$ above which the grooming algorithm does not affect the jet mass anymore and we get back to the ungroomed case. This can be seen as follows. Using $k^+=\omega_J\tau/4$, we rewrite the soft drop criterion in Eq.~(\ref{eq:SD-cs}) as
\be
\f{k^+}{k^-}<\f{R^2}{4}\left(\frac{\tau}{R^2 z_{\rm cut}}\right)^{\frac{2}{2+\beta}} \,.
\label{eq:constrain2-transition}
\ee
We observe that the soft drop constraint is thus less restrictive than the jet algorithm constraint for $\tau/R^2>z_{\rm cut}$ and $\beta>0$: Eq.~\eqref{eq:constrain2-transition} vs.~Eq.~\eqref{eq:constraint2}. In this kinematic region we can therefore remove the $\Theta$-function in Eq.~(\ref{eq:softpass}) associated with the soft drop grooming algorithm and we are left with the soft function for the ungroomed case. Below the transition point the grooming procedure required us to replace the ungroomed soft mode with two modes, $S_i^{\rm gr}$ and $S_i^{\notin {\rm gr}}$. Above the transition point we now find that $S_i^{\rm gr}$ is reduced to the ungroomed soft function which also implies $S_i^{\notin {\rm gr}} \to 1$. In this sense, the entire factorization theorem for the soft drop groomed jet mass in Eq.~(\ref{eq:refactorize-groom}) reduces to the ungroomed case, see Eq.~(\ref{eq:refactorize}), where there is only one (ungroomed) soft function. The result for the ungroomed soft function at NLO was given in Eq.~(\ref{eq:ren_S_ungroom}). In section~\ref{sec:groomed_jet}, the transition point at $\tau/R^2=z_{\rm cut}$ can be seen directly in the numerical studies of the groomed jet mass distribution.

\bibliographystyle{JHEP}
\bibliography{bibliography}

\end{document}